\documentclass[lettersize,journal]{IEEEtran}
\usepackage{amsmath,graphicx}
\usepackage{soul}
\usepackage{pifont}
\usepackage{arydshln}
\usepackage{multirow}
\usepackage{afterpage}
\usepackage{array,colortbl}
\usepackage{xcolor}
\usepackage[numbers,sort&compress]{natbib}
\usepackage[utf8]{inputenc} %
\usepackage[T1]{fontenc}    %
\usepackage{hyperref}       %
\usepackage{url}            %
\usepackage{booktabs}       %
\usepackage{mathtools,amssymb}
\usepackage{amsfonts}       %
\usepackage{nicefrac}       %
\usepackage{microtype}      %
\usepackage{pgfplots,pgfplotstable}
\pgfplotsset{compat=1.14}
\usepackage{array,colortbl}
\usepackage{xcolor}
\usepackage{algorithm,algorithmicx,algpseudocode}
\usepackage[capitalise]{cleveref}
\usepackage{caption}
\usepackage{graphbox}
\usepackage{placeins}
\usepackage{wrapfig}
\usepackage{etoolbox}
\usepackage{subfloat}
\usepackage{graphicx}
\usepackage[thicklines]{cancel}
\usepackage{glossaries-extra}
\usepackage{siunitx}

\newcommand{\bI}{\mathbf{I}}

\newcommand{\bzero}{\mathbf{0}}

\newcommand{\bx}{\mathbf{x}}

\newcommand{\bz}{\mathbf{z}}

\newcommand{\bepsilon}{{\boldsymbol{\epsilon}}}

\usepackage{amsmath}
\usepackage{caption}
\usepackage{subcaption}
\usepackage{bm}

\renewcommand{\vec}[1]{\mathbf{#1}}

\newcommand{\nv}{\vec{n}}

\newcommand{\xv}{\vec{x}}

\newcommand{\zv}{\vec{z}}

\newcommand{\Lc}{{\cal L}}

\newcommand{\LB}{\left(}
\newcommand{\RB}{\right)}

\newcommand{\removed}[1]{}%

\newacronym{ACM}{ACM}{adaptive coding and modulation}
\newacronym{ADC}{ADC}{analog-to-digital conversion}
\newacronym{AGC}{AGC}{automatic gain control}
\newacronym{AWGN}{AWGN}{additive white Gaussian noise}
\newacronym{BER}{BER}{bit error rate}
\newacronym{BLER}{BLER}{block error rate}
\newacronym{BP}{BP}{backpropagation}
\newacronym{BPTT}{BPTT}{backpropagation through time}
\newacronym{CE}{CE}{cross-entropy}
\newacronym{CFO}{CFO}{carrier frequency offset}
\newacronym{CSI}{CSI}{channel state information}
\newacronym{DAC}{DAC}{digital-to-analog conversion}
\newacronym{DL}{DL}{deep learning}
\newacronym{DFT}{DFT}{discrete Fourier transform}
\newacronym{FFT}{FFT}{fast Fourier transform}
\newacronym{GAN}{GAN}{generative adversarial network}
\newacronym{GRU}{GRU}{gated recurrent unit}
\newacronym{iid}{i.i.d.\@}{independent and identically distributed}
\newacronym{IFFT}{IFFT}{inverse fast Fourier transform}
\newacronym{KL}{KL}{Kullback-Leibler}
\newacronym{LSTM}{LSTM}{long short-term memory}
\newacronym{MDP}{MDP}{Markov decision process}
\newacronym{ML}{ML}{machine learning}
\newacronym{MLP}{MLP}{multilayer perceptron}
\newacronym{MIMO}{MIMO}{multiple-input multiple-output}
\newacronym{MSE}{MSE}{mean squared error}
\newacronym{NN}{NN}{neural network}
\newacronym{DNN}{DNN}{deep neural network}
\newacronym{OFDM}{OFDM}{orthogonal frequency-division multiplexing}
\newacronym{pdf}{pdf}{probability density function}
\newacronym{pmf}{pmf}{probability mass function}
\newacronym{PSNR}{PSNR}{peak signal to noise ratio}
\newacronym{RBF}{RBF}{Rayleigh block-fading}
\newacronym{ReLU}{ReLU}{rectified linear unit}
\newacronym{RL}{RL}{reinforcement learning}
\newacronym{RNN}{RNN}{recurrent neural network}
\newacronym{SFO}{SFO}{sampling frequency offset}
\newacronym{SNR}{SNR}{signal-to-noise ratio}
\newacronym{SINR}{SINR}{signal-to-interference-plus-noise ratio}
\newacronym{SGD}{SGD}{stochastic gradient descent}
\newacronym{wrt}{w.r.t.\@}{with respect to}

\newacronym{OAC}{OAC}{over-the-air computation}
\newacronym{MAC}{MAC}{multiple access channel}
\newacronym{SIC}{SIC}{successive interference cancellation}
\newacronym{TDMA}{TDMA}{time division multiple access}
\newacronym{NOMA}{NOMA}{non-orthogonal multiple access}
\newacronym{CL}{CL}{curriculum learning}
\newacronym{JSCC}{JSCC}{joint source-channel coding}
\newacronym{DeepJSCC}{DeepJSCC}{deep joint source-channel coding}
\newacronym{MTL}{MTL}{multi-task learning}
\newacronym{MIL}{MIL}{multi-instance learning}
\newacronym{DML}{DML}{deep metric learning}
\newacronym{IoT}{IoT}{Internet of Things}
\newacronym{SSIM}{SSIM}{structural similarity index measure}
\newacronym{MS-SSIM}{MS-SSIM}{multi-scale \gls{SSIM}}
\newacronym{DDPM}{DDPM}{denoising diffusion probabilistic model}
\newacronym{MVL}{MVL}{multi-view learning}
\newacronym{CNN}{CNN}{convolutional neural network}
\newacronym{LPIPS}{LPIPS}{learned perceptual image patch similarity}
\newacronym{BPG}{BPG}{Better Portable Graphics}

\newacronym{IoE}{IoE}{Internet of Everything}

\newacronym{V2X}{V2X}{vehicle-to-everything}

\newacronym{AR/VR}{AR/VR}{augmented/virtual reality}

\newacronym{DSC}{DSC}{distributed source coding}

\newacronym{ANN}{ANN}{artificial neural network}

\newacronym{BCR}{BCR}{bandwidth compression ratio}
\newacronym{BR}{BR}{bandwidth ratio}
\newacronym{OPTA}{OPTA}{Optimal Performance Theoretically Attainable}
\newacronym{AF}{AF}{attention feature}

\newacronym{DDNM}{DDNM}{deep denoising null-space model}

\newacronym{IR}{IR}{image restoration}
\newacronym{GT}{GT}{ground-truth}
\newacronym{SVD}{SVD}{singular value decomposition}

\begin{document}

\title{SING: Semantic Image Communications using Null-Space and INN-Guided Diffusion Models}

\author{Jiakang Chen\IEEEauthorrefmark{1}\IEEEauthorrefmark{2},~\IEEEmembership{Graduate Student Member,~IEEE}, Selim F. Yilmaz\IEEEauthorrefmark{1},~\IEEEmembership{Member,~IEEE}, \\Di You\IEEEauthorrefmark{1},~\IEEEmembership{Graduate Student Member,~IEEE}, Pier Luigi Dragotti,~\IEEEmembership{Fellow,~IEEE}, and Deniz Gündüz,~\IEEEmembership{Fellow,~IEEE}
\thanks{Jiakang Chen, Selim F. Yilmaz, Di You, Pier Luigi Dragotti, and Deniz Gündüz are with the Department of Electrical and Electronic Engineering, Imperial College London, London SW7 2AZ, U.K. (e-mails: (jiakang.chen21, s.yilmaz21, di.you22, p.dragotti, d.gunduz)@imperial.ac.uk) }%

\thanks{We acknowledge funding from the European Union’s Horizon 2020 Marie Skłodowska Curie Innovative Training Network Greenedge (GA. No. 953775), from the UKRI for the projects INFORMED-AI (EP/Y028732/1), and the SNS JU project 6G-GOALS (Grant Agreement No. 101139232). 

For the purpose of open access, the authors have applied a Creative Commons Attribution (CC BY) license to any Author Accepted Manuscript version arising from this submission.}
\thanks{\IEEEauthorrefmark{1}These authors contributed equally to this work. \IEEEauthorrefmark{2}Corresponding author.}
}

\markboth{Journal of \LaTeX\ Class Files,~Vol.~14, No.~8, August~2021}%
{Shell \MakeLowercase{\textit{et al.}}: A Sample Article Using IEEEtran.cls for IEEE Journals}

\maketitle

\begin{abstract}
Joint source-channel coding systems based on deep neural networks (DeepJSCC) have recently demonstrated remarkable performance in wireless image transmission. Existing methods primarily focus on minimizing distortion between the transmitted image and the reconstructed version at the receiver, often overlooking perceptual quality. This can lead to severe perceptual degradation when transmitting images under extreme conditions, such as low bandwidth compression ratios (BCRs) and low signal-to-noise ratios (SNRs). In this work, we propose SING, a novel two-stage JSCC framework that formulates the recovery of high-quality source images from corrupted reconstructions as an inverse problem. Depending on the availability of information about the DeepJSCC encoder/decoder and the channel at the receiver, SING can either approximate the stochastic degradation as a linear transformation, or leverage invertible neural networks (INNs) for precise modeling. Both approaches enable the seamless integration of diffusion models into the reconstruction process, enhancing perceptual quality. Experimental results demonstrate that SING outperforms DeepJSCC and other approaches, delivering superior perceptual quality even under extremely challenging conditions, including scenarios with significant distribution mismatches between the training and test data.
\end{abstract}


\section{Introduction}
Shannon's separation theorem~\cite{shannon1948mathematical} laid the foundations for modern communication systems. In a typical point-to-point digital communication system, the encoding process at the transmitter involves two steps: source coding followed by channel coding and modulation. The source encoder compresses the input signal by removing possible inherent redundancies, while the channel encoder adds structured redundancy to the compressed bits to correct potential errors caused by the noisy communication channel. For example, in wireless image transmission applications, which will be the main focus of this paper, image compression algorithms such as JPEG or BPG are used to remove redundancies in the input image, and channel coding methods like LDPC or polar codes help to compensate for errors over the channel. Separation theorem states that this two-step procedure is without loss of optimality in the asymptotic infinite blocklength regime for ergodic sources and channels~\cite{vembu1995source}. However, the theoretical optimality of this separate design breaks down in practically relevant finite blocklength regimes as well as in most multi-user setups~\cite{gunduz2009source}. In these scenarios, it is well-known that joint source-channel coding (JSCC) methods that integrate source compression and channel coding into a single step can potentially provide better performance, particularly under extreme bandwidth and latency constraints.

Despite potential gains, designing practical JSCC schemes has long been a challenge~\cite{gunduz2024joint}. Recently, JSCC methods that benefit from modern deep learning (DL) architectures, known as DeepJSCC~\cite{bourtsoulatze2019deep,kurka2020deepjscc,tung2022deepjscc,tung2021deepwive,yang2022ofdm,kurka2021bandwidth,xu2021wireless} have demonstrated remarkable results for a variety of source modalities and channels. DeepJSCC models the encoding and decoding processes of the communication system as an autoencoder, allowing it to extract complex features from the source data and implicitly incorporate the process of combating bandwidth compression and channel noise into its encoding process, thereby significantly improving performance. Additionally, DeepJSCC avoids the cliff effect and achieves smooth degradation, meaning that even if the channel quality is below the target SNR, images can still be decoded within reasonable reconstruction quality. These characteristics give DeepJSCC significant advantages over traditional separation schemes, particularly in challenging channel modeling and estimation scenarios~\cite{xu2023deep}.

Standard DeepJSCC methods typically focus on pixel-level distortion or structural similarity, which fail to account for the semantic similarity between the input and the reconstructed images. It has been recently shown that the perceptual quality of the reconstructed image can be better captured by the learned perceptual image patch similarity (LPIPS)~\cite{zhang2018unreasonable} metric, which exhibits remarkable parallels with human perception. This is crucial in semantic communication, especially in image/video transmission contexts, where the receiver might be more concerned with specific tasks like classification or retrieval rather than precise reconstruction of the original signal. In extreme image compression situations, the receiver might be more interested in generating outputs that share similar semantics as the source signal rather than its precise reconstruction.

In this study, we propose a semantic communication scheme targeting extreme channel environments. Specifically, this scheme aims to improve the perceptual quality of existing DeepJSCC methods under such challenging conditions. We focus on extreme channel settings because, under very low channel signal-to-noise ratios (SNR) and extremely low bandwidth compression ratios (BCRs), DeepJSCC systems optimized by minimizing pixel-level distortion often experience significant degradation in perceptual quality. To address this issue, we aim to incorporate diffusion models~\cite{ho2020denoising}, the state-of-the-art generative models, to achieve high perceptual quality reconstructions.

The method we propose does not rely on a specific implementation of the DeepJSCC encoder/decoder and can enhance the perceptual quality of any JSCC system, achieving fully zero-shot/unsupervised capability. This is because our method directly acts on the output of the original DeepJSCC decoder, treating the restoration of high perceptual quality from the decoder output as an inverse problem, and employing unconditional generative diffusion models. If we have no prior information on the DeepJSCC architecture or the communication channel, we employ the denoising diffusion null-space model (DDNM)~\cite{wang2022zero}, where we approximate the nonlinear degradation process caused by DeepJSCC and the channel as a known linear transformation and construct a pseudo-inverse based on this approximation to guide the pre-trained unconditional diffusion model to restore the degraded output. We call this method SING-Zero as it is a zero-shot scheme oblivious of the specific DeepJSCC architecture. If we have partial information about the DeepJSCC architecture and the underlying channel, we can perform a second stage method, called SING-INN, to further refine the perceptual quality. Inspired by the InDigo architecture in~\cite{you2024indigo+}, we directly model the non-linear degradation process using an invertible neural network (INN) combined with the DDNM method to guide the diffusion process. By employing a more precise degradation model, the SING-INN method further enhances the perceptual quality compared to SING-Zero.

We remark that SING does not require any modification of the underlying physical layer encoding and decoding functions, and can be deployed at the application layer at the receiver end. For SING-Zero, all we need is a powerful pretrained diffusion model for the transmitted images, while SING-INN further requires a small dataset of transmitted and reconstructed image pairs for training together with the underlying communication status.

\subsection{Contributions}
Our main contributions can be summarized as follows:
\begin{enumerate}
\item We propose a two-stage generative restoration method for \gls{DeepJSCC}-based wireless image transmission, accommodating scenarios where prior knowledge of the \gls{DeepJSCC} encoder/decoder functions and the communication channel may or may not be available at the decoder.
\item In the absence of prior information, we present SING-Zero, where we adapt the DDNM framework~\cite{wang2022zero} by approximating the nonlinear degradation induced by \gls{DeepJSCC} and the noisy communication channel as a linear transformation, constructing a pseudo-inverse to guide an unconditional diffusion model for restoring degraded outputs and improving perceptual quality.
\item When partial information about the \gls{DeepJSCC} system and channel is available, we propose SING-INN, where we further refine the restoration process using INN-based modeling inspired by InDigo+~\cite{you2024indigo+}. This approach enhances perceptual quality by accurately capturing non-linear degradations due to communications combined with DDNM guidance.
\item Extensive experiments demonstrate that our method outperforms traditional \gls{DeepJSCC} and recent generative \gls{DeepJSCC} approaches under varying \gls{SNR} and bandwidth conditions, achieving superior generalization without requiring modifications at the transmitter.
\end{enumerate}

The rest of the paper is organized as follows. In \cref{sec:related_work}, we review related works on JSCC and also provide an overview of diffusion models. In \cref{sec:problem_statement}, we define the image transmission problem over noisy channels and introduce key concepts. In \cref{sec:methodology}, we describe our novel methodology, including the use of DeepJSCC, linear approximations, and INN-guided diffusion models. In \cref{sec:experimental_results}, we present our experimental results, showing the performance improvements of our method. Finally, we conclude in \cref{sec:conclusion}.

\section{Related Work}
\label{sec:related_work}
\subsection{Overview of DeepJSCC Methods}
In recent years, JSCC techniques based on deep learning have been extensively studied~\cite{bourtsoulatze2019deep,kurka2020deepjscc,tung2022deepjscc,tung2021deepwive,yang2022ofdm,kurka2021bandwidth,xu2021wireless}. We refer the readers to~\cite{gunduz2024joint} for a recent survey on the topic. In~\cite{bourtsoulatze2019deep}, deep neural networks were first used for wireless image transmission in the context of JSCC. This approach is known as DeepJSCC. DeepJSCC employs an autoencoder architecture for wireless communication, where the encoder converts the source image directly into channel symbols similar to analog communication. Correspondingly, the decoder is used to reconstruct the image directly from the channel symbols corrupted by channel noise. Notably, DeepJSCC has been shown to have smooth degradation with channel quality, avoiding the \textit{cliff effect} while outperforming traditional separation-based digital communication techniques. DeepJSCC has then been extended to incorporate feedback~\cite{kurka2020deepjscc,wu2024transformer} and applied to various communication scenarios such as OFDM~\cite{yang2022ofdm,wu2022channel} and MIMO~\cite{wu2024deep,zhou2024feature}. In addition to wireless image transmission, DeepJSCC has also been extended to video transmission~\cite{tung2021deepwive,zhou2024multi}. However, these widely applicable DeepJSCC methods often focus only on classical pixel-level distortion metrics, neglecting perceptual quality, which is crucial for next-generation semantic communication.

Generative models based on neural networks, such as variational autoencoders~\cite{kingma2013auto} (VAEs), generative adversarial networks~\cite{goodfellow2020generative} (GANs), and diffusion models~\cite{ho2020denoising} have demonstrated powerful capabilities in generating high perceptual quality data. Some studies have attempted to apply these generative models to JSCC. For instance, JSCC schemes have been implemented through VAE~\cite{choi2019neural}, adversarial training for secure JSCC in the presence of eavesdroppers~\cite{marchioro2020adversarial}, GAN-style loss for training decoders~\cite{yang2022ofdm}, and diffusion denoising models to enhance reconstruction quality at the decoder~\cite{niu2023hybrid}. %

In~\cite{bora2017compressed}, pre-trained generative models are used to address the classical inverse problem in compressed sensing. Similarly,~\cite{menon2020pulse} achieves outstanding performance in single-image super-resolution by leveraging generative models, highlighting their effectiveness in solving practical inverse problems. The similarity between communication systems and inverse problems indicates the potential for applying powerful generative models to JSCC, as reflected in recent works such as~\cite{erdemir2022privacy,marchioro2020adversarial,erdemir2022generative,yilmaz2024high}. Notably, the authors of~\cite{erdemir2022generative} proposed InverseJSCC, treating the recovery of high-quality source images from degraded reconstructions as an inverse problem. By utilizing a pre-trained StyleGAN-2 generator and the intermediate latent optimization (ILO) method~\cite{Daras2021IntermediateLO}, they achieve notable perceptual quality results for image reconstruction under extreme channel conditions. In~\cite{tang2024evolving}, the approach in~\cite{erdemir2022generative} is extended to a dynamic generative scenario, where the receiver, rather than relying on a fixed pre-trained generative model, keeps fine-tuning its model using the images it has received in the past. This way the model continuously adapts to the statistics of the images being transmitted.

Diffusion models have made significant advances in the field of visual generation. As a result, some studies have attempted to leverage powerful diffusion models to solve inverse problems, particularly in image restoration tasks. Among these, we are particularly interested in zero-shot methods because they allow us to directly apply diffusion models, trained in an unconditioned manner, to image restoration tasks without the need for supervised tuning of the diffusion model to a specific degradation. For example,~\cite{choi2021ilvr} guides the diffusion model outputs to restored images by replacing the low-frequency components in the denoised output with those from the reference degraded image. In~\cite{chung2023diffusion}, the authors further extend the solvers of diffusion models through approximate posterior sampling, effectively addressing the problem of general noisy image restoration.

Due to the powerful generative capabilities of diffusion models, they have also been applied in semantic communications. In~\cite{niu2023hybrid}, the authors propose a hybrid JSCC scheme that combines traditional transmission methods with a process of adding noise to input images through diffusion and then transmitting them via DeepJSCC, achieving high-quality reconstruction at the decoder. In~\cite{wu2024cddm}, the authors apply the noise-removal properties of diffusion models to wireless communication to help the receiver mitigate the impact of channel noise. In~\cite{guo2024diffusion}, the authors consider the application of diffusion models in wireless communication under strict bandwidth constraints. In this approach, the signal transmission process through the channel is treated as the forward process in a diffusion model and a VAE-based upsampling and downsampling pair is used to reduce the bandwidth requirements by leveraging the advanced compression capabilities of VAEs. In \cite{zhang2025semantics}, the authors propose conditional diffusion using some semantic description of the image (e.g., text or edge map), and employ a semantics-guided channel denoising approach allowing dynamic adaptation to channel conditions.

In \cite{chen2024commin}, the authors propose CommIN, a JSCC scheme that treats image reconstruction as an inverse problem, using INNs and diffusion models to enhance perceptual quality, outperforming DeepJSCC in extreme conditions. Their INN-based degradation modeling effectively separates coarse and detail components, allowing diffusion models to refine reconstructions. Similarly, in \cite{yilmaz2024high}, the authors introduce a DeepJSCC framework with denoising diffusion models, transmitting lower-resolution images and restoring details at the receiver, achieving superior perceptual quality. By leveraging range-null space decomposition, their method reconstructs high-fidelity images with notable improvements in perception- and distortion-oriented metrics. Our work is different from \cite{zhang2025semantics,niu2023hybrid,wu2024cddm,guo2024diffusion} in that, our approach can be designed in conjunction with any JSCC scheme, and does not require any modification of the underlying channel coding and modulation modules.

\subsection{A Brief Review of Denoising Diffusion Probabilistic Models}
Denoising diffusion probabilistic models (DDPMs)~\cite{ho2020denoising} represent a powerful class of generative models that excel at learning complex data distributions by progressively refining noisy inputs. These models operate through two complementary processes: a forward diffusion process, which gradually corrupts data by adding noise, and a reverse denoising process, which reconstructs the original data by removing the noise in a step-by-step manner.

The forward process in DDPM is designed to transform data into a tractable distribution, typically a Gaussian distribution. This transformation occurs over $T$ discrete time steps. Starting from an initial data sample $\mathbf{x}_0$ from the target distribution, at each time step $t$, Gaussian noise is incrementally added to produce the noisy state $\mathbf{x}_t$. The process is defined by:
\begin{equation}
    q(\mathbf{x}_t | \mathbf{x}_{t-1}) = \mathcal{N}(\mathbf{x}_t; \sqrt{1-\beta_t}\mathbf{x}_{t-1}, \beta_t\mathbf{I}),
    \label{eq:ddpm_forward}
\end{equation}
where $\beta_t$ is a variance schedule that controls the noise level at each time step. Over the sequence of $T$ steps, the data distribution is gradually morphed into a simple Gaussian distribution. Remarkably, any intermediate state $\mathbf{x}_t$ can be directly sampled from the original data $\mathbf{x}_0$ using the expression:
\begin{equation}
    \mathbf{x}_t = \sqrt{\bar{\alpha}_t}\mathbf{x}_0 + \sqrt{1-\bar{\alpha}_t} \boldsymbol{\epsilon},
    \label{eq:xt}
\end{equation}
where $\alpha_t = 1 - \beta_t$, $\bar{\alpha}_t = \prod_{i=1}^t \alpha_i$ and $\boldsymbol{\epsilon} \sim \mathcal{N}(0, \mathbf{I})$ is a Gaussian noise term. This direct sampling formula is central to the forward process as it simplifies the generation of noisy data at any time step.

The reverse process is where the generative power of DDPM is realized. It aims to invert the forward process, starting from Gaussian noise and iteratively denoising it back to the original data distribution. This reverse process is modeled as a learned sequence of denoising steps. The conditional distribution of each reverse step can be computed using Bayes' theorem:
\begin{equation}
    \mathbf{x}_{t-1} = \frac{1}{\sqrt{\alpha_t}} \left( \mathbf{x}_t - \frac{1-\alpha_t}{\sqrt{1-\bar{\alpha}_t}} \boldsymbol{\epsilon} \right) + \sigma_t \mathbf{z},
    \label{eq:ddpm_reverse}
\end{equation}
where $\sigma_t = \sqrt{\frac{1-\bar{\alpha}_{t-1}}{1-\bar{\alpha}_t}\beta_t}$, and $\mathbf{z} \sim \mathcal{N}(0, \mathbf{I})$ represents additional Gaussian noise, ensuring that the reverse step maintains the stochasticity necessary for generative diversity.

To predict the noise term $\boldsymbol{\epsilon}$ at each time step, DDPM employs a neural network $\boldsymbol{\epsilon}_\theta(\mathbf{x}_t, t)$. The training objective of this network is to minimize the difference between the true noise $\boldsymbol{\epsilon}$ and the predicted noise $\boldsymbol{\epsilon}_\theta(\mathbf{x}_t, t)$, given by:
\begin{equation}
    \nabla_{\boldsymbol{\theta}} || \boldsymbol{\epsilon} - \boldsymbol{\epsilon}_\theta(\sqrt{\bar{\alpha}_t}\mathbf{x}_0 + \sqrt{1-\bar{\alpha}_t} \boldsymbol{\epsilon}, t) ||^2_2,
    \label{eq:ddpm_loss}
\end{equation}
where $\mathbf{x}_0$ represents the original data, and $\boldsymbol{\epsilon} \sim \mathcal{N}(0, \mathbf{I})$ is the noise added during the forward process.

The generative process in DDPM is executed by iteratively applying the reverse denoising steps, starting from an initial noise sample $\mathbf{x}_T \sim \mathcal{N}(\mathbf{0}, \mathbf{I})$. As the model progresses through each time step, it gradually refines the noise into a data sample resembling those in the training distribution. This iterative denoising effectively transforms random noise into a meaningful data sample, thereby showcasing the model's ability to learn and reproduce complex data distributions.
\section{Problem Statement}  
\label{sec:problem_statement}

\begin{figure}[t]
    \centering
    \includegraphics[width=\columnwidth]{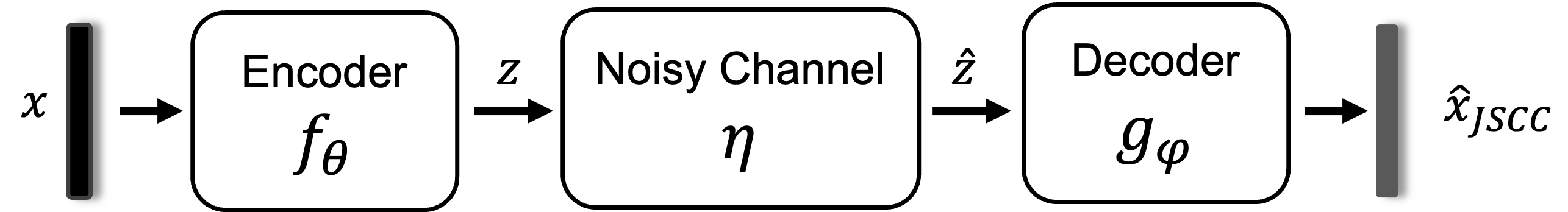}
\caption{Overview of JSCC-based communication systems. }
    \label{fig:jsccsystem}
\end{figure}

\begin{figure*}[t!]
\centering
\includegraphics[width=0.8\linewidth]{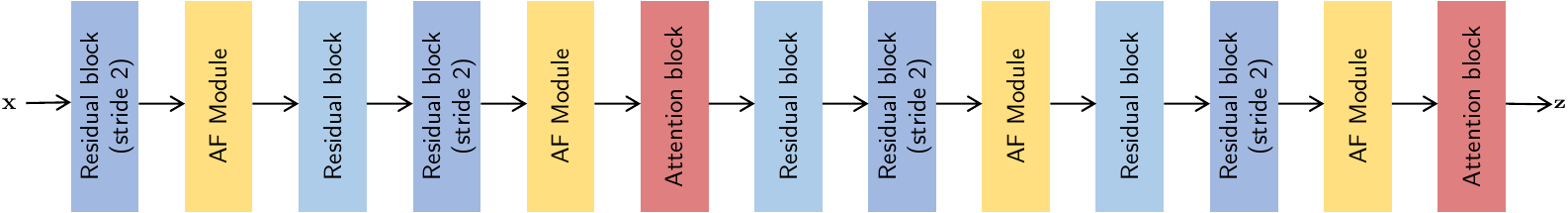}
\vfill
\vspace{0.5em}
\includegraphics[width=0.9\linewidth]{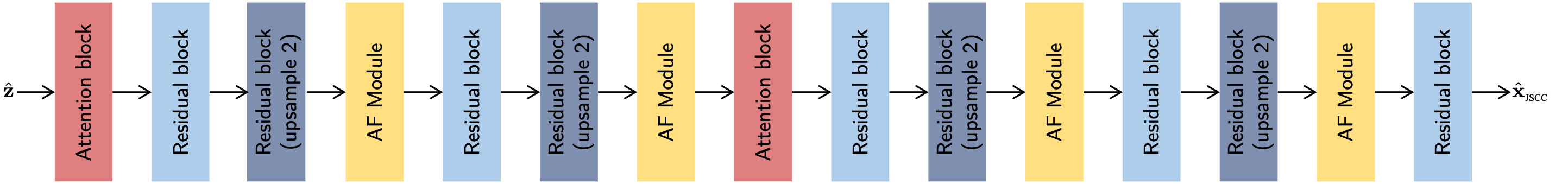}
\caption{Encoder architecture of our method (\textbf{top}), which is used to encode the input image $\xv$ at the transmitter. Decoder architecture of our method (\textbf{bottom}), which is used to reconstruct the input image from the noisy channel output $\hat{\zv}$.}
\label{fig:deepjscc_architecture}
\end{figure*}

\label{sec:ddnm}
\begin{figure*}[htbp!]
    \centering
    \includegraphics[scale=0.7]{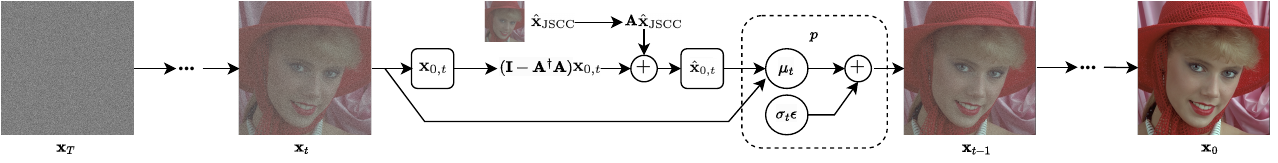}
    \caption{Overview of the SING-Zero with null-space-based image restoration procedure.}
    \label{fig:restoration}
\end{figure*}

We consider point-to-point wireless transmission of images over a noisy channel (see Fig. \ref{fig:jsccsystem}). As in~\cite{bourtsoulatze2019deep}, we model the transmitter and receiver as two neural networks. Specifically, the transmitter is the encoder and the receiver is the decoder of an autoencoder system. The transmitter maps the source signal, a real vector $\mathbf{x} \in \mathbb{R}^m$ to the complex channel input vector $\mathbf{z} \in \mathbb{C}^k$ using a parameterised encoding function $\mathbf{z}=f\left(\mathbf{x} ; \boldsymbol{\theta}\right)$. Here $m$ represents the source bandwidth and $k$ the channel bandwidth. We define the bandwidth compression ratio (BCR) as $\rho \triangleq k/m$ channel uses per source sample. This ratio reflects the compression applied to the signal by the communication system. We expect the performance to improve as more channel bandwidth becomes available, i.e., as $\rho$ increases.
In the image transmission problem, $m=H \times W \times C$, where $H$, $W$ and $C$ are the height, width and, number of colour channels of the input image, respectively. The channel input vector must satisfy the following average power constraint:

\begin{equation}
\frac{1}{k} \mathbb{E}_{\mathbf{z}}\left[\|\mathbf{z}\|_2^2\right] \leq \bar{P}.
\label{eq:power_constraint}
\end{equation}
This is achieved by normalising the signal $\tilde{\mathbf{z}}$, output of the last layer of the encoder, as follows:
\begin{equation}
\mathbf{z}=\sqrt{k \bar{P}} \frac{\tilde{\mathbf{z}}}{\sqrt{{\tilde{\mathbf{z}}}^H {\tilde{\mathbf{z}}}}},
\end{equation}
where $H$ is the Hermitian transpose. The channel introduces random corruptions to the transmitted symbols. We consider a complex AWGN channel so that the received signal can be expressed as follows:
\begin{equation}
\hat{\mathbf{z}}=\eta\left(\mathbf{z}, \sigma^2\right)=\mathbf{z}+\mathbf{n}_C,
\end{equation}
where $\mathbf{n}_C$ is sampled in an independent identically distributed (i.i.d.) way from a complex Gaussian distribution with variance $\sigma^2$: $\mathbf{n}_C\sim\mathcal{C N}\left(0, \sigma^2 \mathbf{I}\right)$. The receiver uses a parameterised decoding function $\hat{\mathbf{x}}_{\mathrm{JSCC}}=g\left(\hat{\mathbf{z}} ; \boldsymbol{\phi}\right)$ to get the degraded reconstruction $\hat{\mathbf{x}}_{\mathrm{JSCC}}$ from $\hat{\mathbf{z}}$. To represent the condition of the noisy channel, we define channel SNR as follows:
\begin{equation}
\text{SNR} \triangleq 10 \text{log}_{10} \frac{\bar{P}}{\sigma^2} \mathrm{~dB}.
\end{equation}
The encoder and decoder are jointly optimised by minimising:
\begin{equation}
\arg \min _{\boldsymbol{\theta}, \boldsymbol{\phi}} \mathbb{E}_{\mathbf{x} \sim p_{\mathbf{x}}} \mathbb{E}_{\hat{\mathbf{x}} \sim p_{\hat{\mathbf{x}} \mid \mathbf{x}}}[d(\mathbf{x}, \hat{\mathbf{x}}_{\mathrm{JSCC}})],
\end{equation} 
where $d$ can be any distortion measure.

In our simulations, we will employ the same DeepJSCC architecture as in~\cite{erdemir2022generative}. 
We note that DeepJSCC models can be optimized directly for any differentiable distortion metric and can be readily adapted for a range of multimedia transmission tasks, such as video transmission as demonstrated in~\cite{tung2021deepwive}. Following the methodology described in~\cite{mentzer2020high}, we adopt a composite loss function, $\Lc$, to simultaneously optimize the \emph{distortion-perception trade-off}~\cite{erdemir2022generative} for the reconstructed images, formulated as: \begin{equation} 
\Lc \LB \hat{\mathbf{x}}_{\mathrm{JSCC}}, \mathbf{x} \RB = \operatorname{MSE}(\hat{\mathbf{x}}_{\mathrm{JSCC}}, \mathbf{x}) + \lambda \cdot \operatorname{LPIPS}(\hat{\mathbf{x}}_{\mathrm{JSCC}}, \mathbf{x}), \label{eq } 
\end{equation} 
where \gls{LPIPS}~\cite{zhang2018unreasonable} is employed as a perceptual quality metric. The parameter $\lambda$ regulates the balance between the \gls{LPIPS} and \gls{MSE} losses. Designed to assess the similarity of distorted patches, \gls{LPIPS} calculates the distance in the feature space of a deep neural network (DNN) model initially trained for image classification. In contrast to the \gls{MSE} distortion metric which operates on a point-wise basis, perception-oriented distortion metrics like LPIPS~\cite{zhang2018unreasonable} and \gls{MS-SSIM}~\cite{wang2003multiscale} evaluate the similarity between the original and reconstructed images using proxies to human perception.

In the following section, we introduce our JSCC schemes SING-Zero and SING-INN, which leverage diffusion models at the receiver to effectively reduce the semantic distortion between the source signal and its reconstruction, in order to improve the perceptual quality.

\section{Methodology}
\label{sec:methodology}

\subsection{DeepJSCC as an Inverse Problem}
We aim to improve the perceptual quality of the images reconstructed by the DeepJSCC encoder-decoder pair by formulating the reconstruction process as an inverse problem. Specifically, we interpret the received degraded signal $\hat{\mathbf{x}}_{\mathrm{JSCC}}$ as degraded measurements $\mathbf{y}=A(\mathbf{x})$ in a conventional inverse problem, where the forward operator $A$ encapsulates the entire DeepJSCC communication chain, including the encoder, noisy channel, and decoder. Formally, $A$ is expressed as:
\begin{equation}
\mathbf{y}=A(\mathbf{x}, \eta)
= g_\phi\left(\eta\left(f_\theta(\mathbf{x}), \sigma^2\right)\right)
=\hat{\mathbf{x}}_{\mathrm{JSCC}},
\label{eq:eq11}
\end{equation}
representing a non-linear stochastic process with AWGN, $\eta$. \Cref{fig:deepjscc_architecture} depicts the employed DeepJSCC encoder and decoder architectures, which can be substituted with other DeepJSCC-based or traditional source/channel coding techniques.

 With the forward model in place and given measurements of $\mathbf{y}$, one typically seeks to restore the original signal $\mathbf{x}$ by simultaneously imposing a fidelity constraint and a proper prior on $\mathbf{x}$. The fidelity constraint is the MSE between the output of $A$ and the measurements $\mathbf{y}$:
\begin{equation}
\operatorname{MSE}(A(\mathbf{x}), \mathbf{y})=\|A(\mathbf{x})-\mathbf{y}\|_2^2.
\end{equation}
For the prior, we adopt diffusion models, which offer a highly effective way to model the underlying structure of $\mathbf{x}$.%

This approach involves two main challenges: (1) the non-linear nature of the degradation process, which lacks a closed-form expression, and (2) the computational complexity of incorporating a diffusion model as a prior. To tackle these challenges, we propose a two stage scheme depending on the availability of prior knowledge of the coding system and the channel. When no information about the DeepJSCC architecture or communication channel is available, we approximate the degradation process as a known linear transformation. When partial information about the DeepJSCC system and channel is available—evidenced by the received degraded images—we directly model the non-linear degradation process using INNs. This approach allows us to capture and leverage the non-linear properties of the system more effectively. In both cases, we utilize pre-trained unconditional diffusion models to significantly reduce computational complexity while maintaining reconstruction quality.

\subsection{SING-Zero: Approximating the Forward Process as a Linear Transform for Fully Blind Semantic Communications}
Inspired by DDNM~\cite{wang2022zero,yilmaz2024high}, we approximate the forward process $A(\mathbf{x}, \eta)$ as a known linear transformation, denoted by $\mathbf{A} \in \mathbb{R}^{d \times m}$. 
The approximation above can be expressed as
\begin{equation}
\hat{\mathbf{x}}_{\mathrm{JSCC}}\approx\mathbf{A} \mathbf{x}.
\label{eq:eq13}
\end{equation}
The degradation matrix $\mathbf{A}$ can be any linear operator, such as decolorization or mean-pooling-based downsampling~\cite{wang2022zero}. 
An ideal $\mathbf{A}$ should reduce the amount of transmitted information so that $\hat{\mathbf{x}}_{\mathrm{JSCC}}$ can be effectively recovered at the receiver with lower reconstruction error, while also preserving perceptual quality, enabling the restoration of $\hat{\mathbf{x}}_{\mathrm{JSCC}}$ into a consistent and high-quality image $\hat{\mathbf{x}}$.

\begin{algorithm}[t]
  \caption{SING-Zero: Null Space Sampling} \label{alg:sampling_ddnm}
  \small
  \centering
  \begin{algorithmic}[1]
    \vspace{.04in}
    \Require The received degraded image $\hat{\mathbf{x}}_{\mathrm{JSCC}}$, degradation operator $\mathbf{A}$, and its pseudo-inverse $\mathbf{A}^{\dagger}$.
    \State $\bx_T \sim \mathcal{N}(\bzero, \bI)$
    \For{$t=T, \dotsc, 1$}
      \State $\bz \sim \mathcal{N}(\bzero, \bI)$ if $t > 1$, else $\bz = \bzero$

      \State{{${\color{black}{\bx_{0,t}  = \frac{1}{\sqrt{\bar\alpha_t}}(\bx_{t} - \sqrt{1 - \bar\alpha_t} \bepsilon_\theta(\bx_t, t) )}}$}}
      
      \State $\hat{\mathbf{x}}_{0, t}=\mathbf{x}_{0 , t}-\mathbf{A}^{\dagger}\left(\mathbf{A} \mathbf{x}_{0 , t}-\hat{\mathbf{x}}_{\mathrm{JSCC}}\right)$
      
      \State{${\bx}_{t-1}  = \frac{\sqrt{\alpha_t}(1-\bar\alpha_{t-1})}{1 - \bar\alpha_t}\bx_{t}+\frac{\sqrt{\bar\alpha_{t-1}}\beta_t}{1 - \bar\alpha_t}\hat{\bx}_{0,t}  + \sigma_t \bz$}

    \EndFor
    \State \textbf{return} $\bx_0$
    \vspace{.04in}
\end{algorithmic}
  \label{alg:ddnm_sampling}
\end{algorithm}
Given $\mathbf{A}$, we denote with $\mathbf{A}^{\dagger}$ its pseudo-inverse. For example, $\mathbf{A}^{\dagger}$ is the upsampling matrix for a downsampling-based degradation operator $\mathbf{A}$. Given a received degraded image $\hat{\mathbf{x}}_{\mathrm{JSCC}}$, the image restoration aims to produce $\hat{\mathbf{x}}$ with minimal distortion while enhancing perceptual quality. To achieve this, we impose a consistency constraint $\mathbf{A} \hat{\mathbf{x}} \equiv \hat{\mathbf{x}}_{\mathrm{JSCC}}$ and ensure the perceptual quality by requiring $\hat{\mathbf{x}}$ to follow the distribution of ground truth images. We employ the zero-shot image restoration method DDNM to achieve that.

We use the neural network in DDPM, denoted by $\boldsymbol{\epsilon}_\theta$, trained via a progressive denoising task with T-step forward and backward processes. The forward process adds noise to the data, while the reverse process generates samples from this noise. At timestep $t$, the noise in $\mathbf{x}_t$ is predicted by the neural network $\boldsymbol{\epsilon}_\theta$ as $\epsilon_t=\boldsymbol{\epsilon}_\theta\left(\mathbf{x}_t, t\right)$. DDNM uses and guides the pretrained DDPM $\boldsymbol{\epsilon}_\theta$ to improve image quality and consistency without additional training~\cite{wang2022zero}.

Specifically, the sample $\mathbf{x}$ is decomposed using the range-null space decomposition: $\mathbf{x} \equiv \mathbf{A}^{\dagger} \mathbf{A} \mathbf{x}+(\mathbf{I}-$ $\left.\mathbf{A}^{\dagger} \mathbf{A}\right) \mathbf{x}$. Here, $\mathbf{A}^{\dagger} \mathbf{A} \mathbf{x}$ is the range component and $(\mathbf{I}-$ $\left.\mathbf{A}^{\dagger} \mathbf{A}\right) \mathbf{x}$ is the null space component. Let $\mathbf{x}_{0,t}$ denote the estimated $\mathbf{x}_0$ at diffusion timestep $t$. For a received degraded image $\hat{\mathbf{x}}_{\mathrm{JSCC}}$, we form a solution for $\hat{\mathbf{x}}$ that fulfills the consistency constraint:
$\hat{\mathbf{x}}=\mathbf{A}^{\dagger} \hat{\mathbf{x}}_{\mathrm{JSCC}}+\left(\mathbf{I}-\mathbf{A}^{\dagger} \mathbf{A}\right) \mathbf{x}_{0 , t}$,
where $\mathbf{x}_{0,t}$ is refined iteratively using the method shown in \cref{alg:ddnm_sampling}. The hyperparameter $\Bar{\alpha}$ defines the noise schedule. This process ensures that modifying $\mathbf{x}_{0,t}$ does not break the consistency constraint, allowing us to enhance perceptual quality while preserving the transmitted image's consistency with DeepJSCC decoder's output. Finally, $\hat{\mathbf{x}}$ is unflattened back to its original dimensions of $H \times W\times C$.

\subsection{SING-INN: Modelling the Forward Process using INNs}
\begin{figure}[t]
    \centering
    \includegraphics[width=0.46\textwidth]{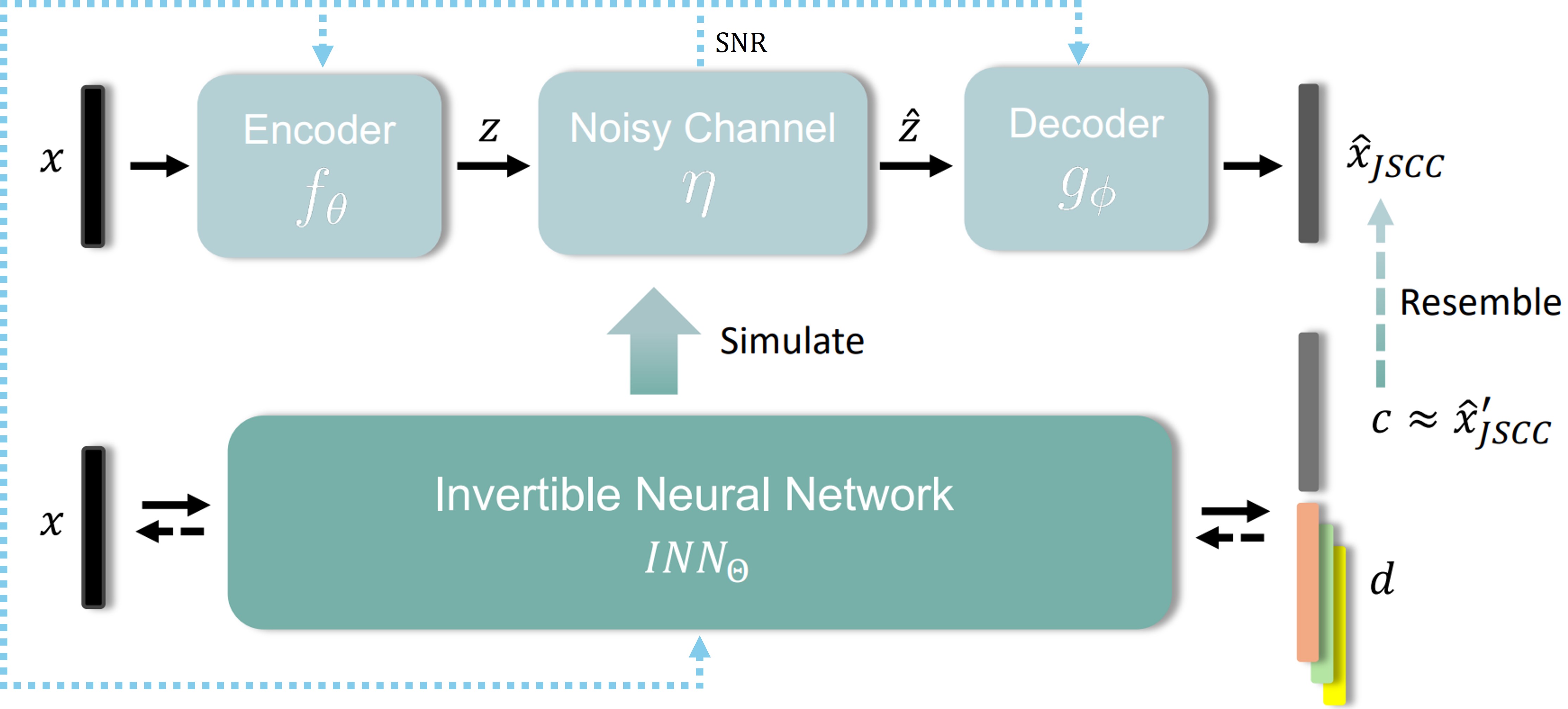}
\caption{The key insight of SING-INN is to simulate the degradation process of the communication pipeline (top) with a conditional INN (bottom), which can transform the input original signal $\mathbf{x}$  into the simulated degraded observation $\hat{\mathbf{x}}'_{\mathrm{JSCC}}$ and estimated lost details $\mathbf{d}$ under a certain channel SNR. At testing stage, $\mathbf{d}$ is estimated using a diffusion model.}
    \label{fig:idea}
\end{figure}
While DDNM-based approach is effective for semantic communication, the assumption that $\mathbf{A}$ is linear limits its performance in communication settings, as the JSCC encoder/decoder operations are typically highly non-linear. Inspired by our recent work CommIN~\cite{chen2024commin}, we further improve our approach using INNs and generative diffusion models. The key insight here is that INN splits the signal (typically an image) into a coarse version $\mathbf{c}$ and details $\mathbf{d}$. In~\cite{you2024indigo+}, the forward part of the INN is trained to mimic the degradation process in an inverse problem so that $\mathbf{c}$ is very close to the degraded image. In this paper, instead, we treat the communication process composed of the transmission, corruption with channel noise and reconstruction as the degradation process and mimic it using an INN (see Fig. \ref{fig:idea}). 

Specifically, building upon the CommIN framework \cite{chen2024commin}, we introduce a conditional INN, where the channel SNR serves as the conditioning input. However, contrary to CommIN, this design allows a single INN to model the degradation process across various SNRs in a unified manner, thereby eliminating the need to train separate models for each channel condition. Consequently, our method extends the advantages of CommIN—effective modeling of the communication degradation process—while significantly enhancing its usability and efficiency in real-world scenarios.

As illustrated in Fig.\ref{fig:idea}, we train a conditional INN, where the channel SNR is used as a conditioning input, to simulate the degradation process. Under a given channel SNR, the INN decomposes the original signal $\mathbf{x}$ into the estimated degraded observation $\hat{\mathbf{x}}'_{\mathrm{JSCC}}$ ($\approx \hat{\mathbf{x}}_{\mathrm{JSCC}}$) and the lost details $\mathbf{d}$. Due to the invertibility of the INN, under the same channel SNR, it is possible to perfectly recover the original image $\mathbf{x}$ if one has access to both $\hat{\mathbf{x}}'_{\mathrm{JSCC}}$ and $\mathbf{d}$. By incorporating the SNR as a condition, a single INN can be used to handle varying channel conditions in a unified manner, thereby simplifying the training and inference processes. During inference, the pre-trained conditional INN guides the diffusion process, ensuring integration of observed measurements $\hat{\mathbf{x}}_{\mathrm{JSCC}}$ and preservation of intricate details.

\begin{algorithm}[t]
  \caption{SING-INN: Null Space Sampling with INN} \label{alg:sampling_inn}
  \small
  \centering
  \begin{algorithmic}[1]
    \vspace{.04in}
    \Require The received degraded image $\hat{\mathbf{x}}_{\mathrm{JSCC}}$, the degradation operator $\mathbf{A}$, and its pseudo-inverse $\mathbf{A}^{\dagger}$, and the pre-trained INN, and the certain channel SNR.
    \State $\bx_T \sim \mathcal{N}(\bzero, \bI)$
    \For{$t=T, \dotsc, 1$}
      \State $\bz \sim \mathcal{N}(\bzero, \bI)$ if $t > 1$, else $\bz = \bzero$

      \State{{${\color{black}{\bx_{0,t}  = \frac{1}{\sqrt{\bar\alpha_t}}(\bx_{t} - \sqrt{1 - \bar\alpha_t} \bepsilon_\theta(\bx_t, t) )}}$}}
      
      \State {$\hat{\mathbf{x}}_{0 , t}=\mathbf{x}_{0 , t}-\mathbf{A}^{\dagger}\left(\mathbf{A} \mathbf{x}_{0 , t}-\hat{\mathbf{x}}_{\mathrm{JSCC}}\right)$}
      
      \State{$\hat{\bx}_{t-1}  = \frac{\sqrt{\alpha_t}(1-\bar\alpha_{t-1})}{1 - \bar\alpha_t}\bx_{t}+\frac{\sqrt{\bar\alpha_{t-1}}\beta_t}{1 - \bar\alpha_t}\hat{\bx}_{0,t}  + \sigma_t \bz$}
      
      \State {\color{blue}$\mathbf{c}_{t},\mathbf{d}_{t}= \textit{INN}_{Forward}(\hat{\bx}_{0,t}, \text{SNR})$}
      
      \State {\color{blue}$\tilde\bx_{0,t}=\textit{INN}_{Inverse}(\hat{\mathbf{x}}_{\mathrm{JSCC}},\mathbf{d}_{t}, \text{SNR})$}  
      
      \State{\color{blue}{$\bx_{t-1} =\hat{\bx}_{t-1}  - { {\zeta}}\nabla_{\bx_{t}} \|{\tilde\bx_{0,t} - \hat{\bx}_{0,t}}\|_2^2$}}
    \EndFor
    \State \textbf{return} $\bx_0$
    \vspace{.04in}
\end{algorithmic}
  \label{1}
\end{algorithm}
\begin{figure}[t]
    \centering
    \includegraphics[width=0.5\textwidth]{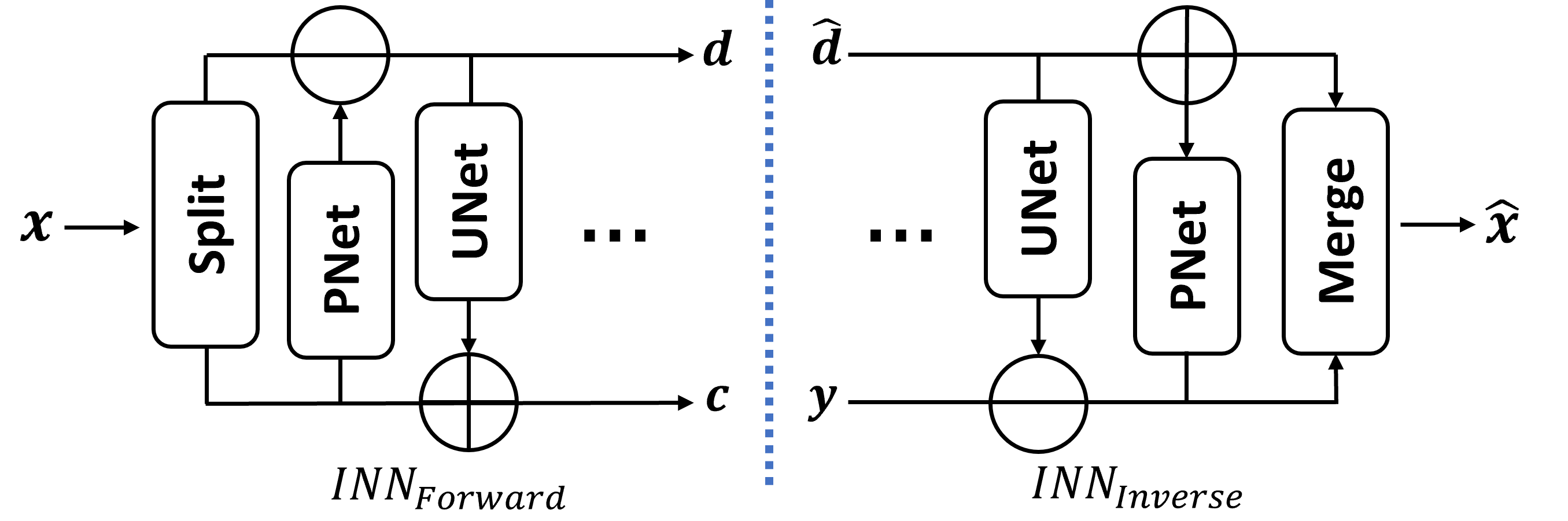}
\caption{General architecture of conditional INN in this work.}  \vspace{-0.25cm}
    \label{fig:Details of our INN}
\end{figure}

The basic architecture of our conditional INN is shown in Fig.~\ref{fig:Details of our INN}. The original image is split into two parts by a splitting operator. Then, the Prediction Network (PNet), conditioned on the coarse part and channel SNR, aims to predict the detail part. Meanwhile, the Update Network (UNet), conditioned on the detail part and channel SNR, is used to adjust the coarse part to make it smoother. The PNet and UNet are applied alternatively to generate the coarse and detail parts, $\mathbf{c}$ and $\mathbf{d}$, respectively. Notably, PNet and UNet can be any complex non-linear functions without compromising the invertibility of the INN. In our implementation, each PNet/UNet is composed of $j$ CondBlocks, each of which can be described as
\begin{align}
\alpha &= \mathrm{FC}(\mathrm{SNR}), \\
x_{j+1} &= \mathrm{Conv}_2\Bigl(\mathrm{ReLU}\bigl(\mathrm{Conv}_1(x_{j})\bigr)\Bigr) \,\odot \alpha + x_{j},
\end{align}
where $\odot$ indicates channel-wise multiplication.

To model the degradation process,  we impose that $\mathbf{c}$ resembles $\hat{\mathbf{x}}_{\mathrm{JSCC}}$. 
Given a training set $\left \{ \mathbf{x}^{i}, {\hat{\mathbf{x}}^{i}_{\mathrm{JSCC}}}\right \}_{i=1}^{N}$, which contains $N$ high-quality images and their low-quality counterparts recovered by DeepJSCC, we optimize our INN with the following loss function: 
\vspace{-0.3cm}
\begin{align}
\begin{split}
L\left ( \Theta  \right )=\frac{1}{N}\sum_{i=1}^{N}\left \| \mathbf{c}^{i}-{\hat{\mathbf{x}}^{i}_{\mathrm{JSCC}}} \right \|_{2}^{2},
\end{split}
\end{align} 
where $\Theta$ denotes the learnable parameters of our INN.
Once we constrain one part of the output of $\textit{INN}_{Forward}(\mathbf{x})$ to be close to $\hat{\mathbf{x}}_{\mathrm{JSCC}}$, due to invertibility, the other part of the output, $\mathbf{d}$, will inevitably represent the detailed information lost during the degradation process. We note that to train our INN model, we do not need to have access to DeepJSCC architecture but only to a set of original and degraded image pairs.

In Algorithm \ref{alg:sampling_inn}, we show how DDNM and INN are combined to improve the quality of the received image $\hat{\mathbf{x}}_{\mathrm{JSCC}}$. Specifically, given $\hat{\mathbf{x}}_{\mathrm{JSCC}}$ as defined in (\ref{eq:eq11}), we first apply the main steps of DDNM in line 4-6, then, given $\hat{\mathbf{x}}_{0 , t}$ which is a preliminary estimate consistent with $\hat{\mathbf{x}}_{\mathrm{JSCC}}$ under a linear degradation model, we improve the consistency by using INN in steps 7-9 (in blue). In these additional steps, given a channel SNR, we decompose the intermediate result $\hat{\mathbf{x}}_{0,t}$ with $\textit{INN}_{Forward}$ into coarse $\mathbf{c}_{t}$ and detail part $\mathbf{d}_{t}$, and replace the coarse part $\mathbf{c}_{t}$ with the received degraded image $\hat{\mathbf{x}}_{\mathrm{JSCC}}$. The INN-optimized $\tilde\bx_{0,t}$ is generated by the inverse process $\textit{INN}_{Inverse}$. Thus, the INN-optimized $\tilde\bx_{0,t}$ is composed of the coarse information $\hat{\mathbf{x}}_{\mathrm{JSCC}}$ and the details generated by the diffusion process. To incorporate the INN-optimized $\tilde\bx_{0,t}$ into the DDNM algorithm, 
we update $\bx_{t}$ with the guidance of the gradient of $\|{\tilde\bx_{0,t} - \hat\bx_{0,t}}\|_2^2$ as shown in step 9. With the help of INN, our algorithm effectively estimates the details lost in the non-linear degradation of the communication process and does not require the knowledge of the closed-form expression of the degradation model.

\begin{minipage}[t]{0.44\textwidth}
    \vspace{-0.45cm}

\end{minipage}

\begin{figure}[!]\footnotesize %
\centering
\hspace{-0.26cm}
\begin{tabular}{c@{\extracolsep{0em}}c@{\extracolsep{0em}}c}
       
		\includegraphics[width=0.35\textwidth]{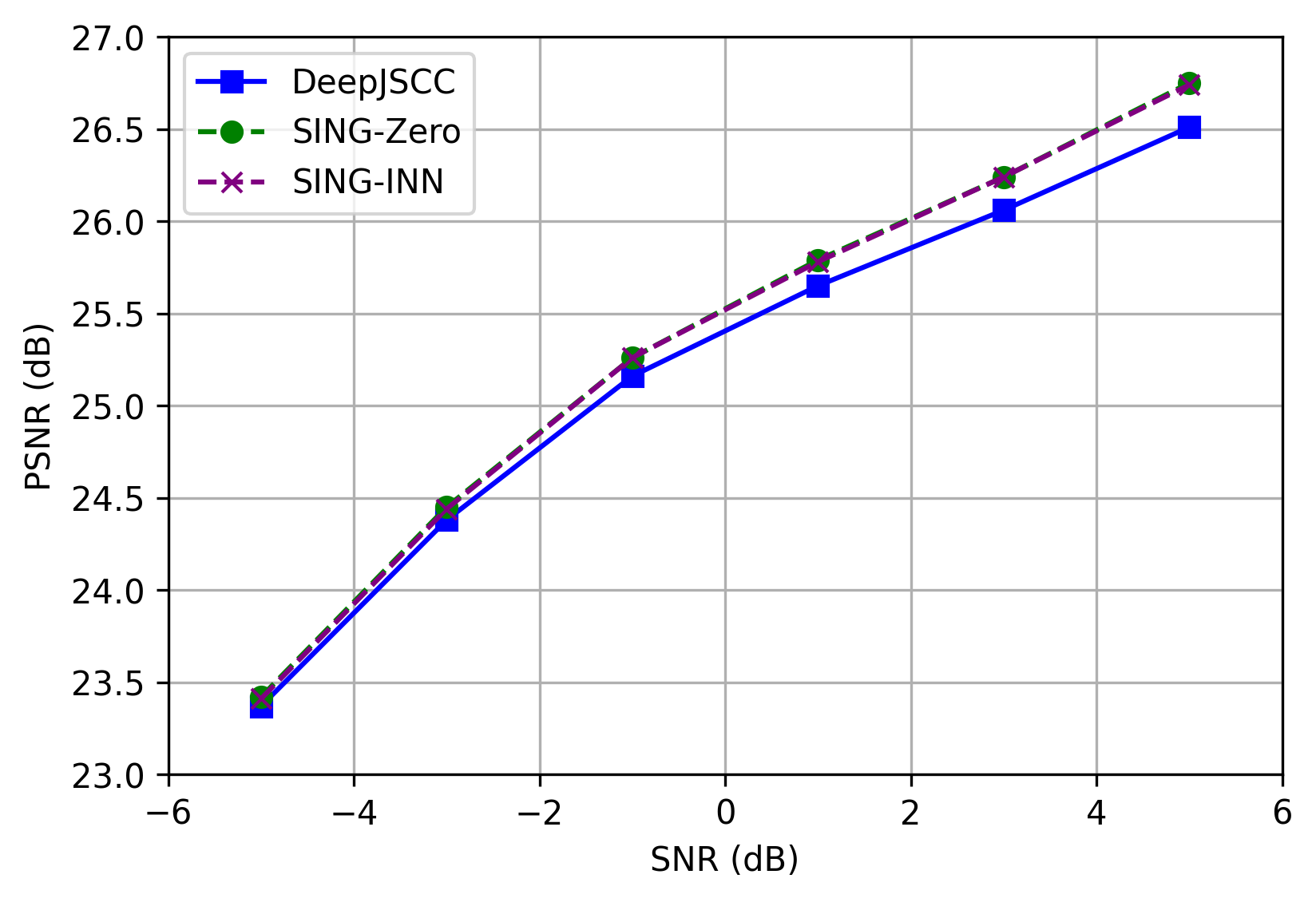}\\
  (a) PSNR versus SNR (higher better)\\
		\includegraphics[width=0.35\textwidth]{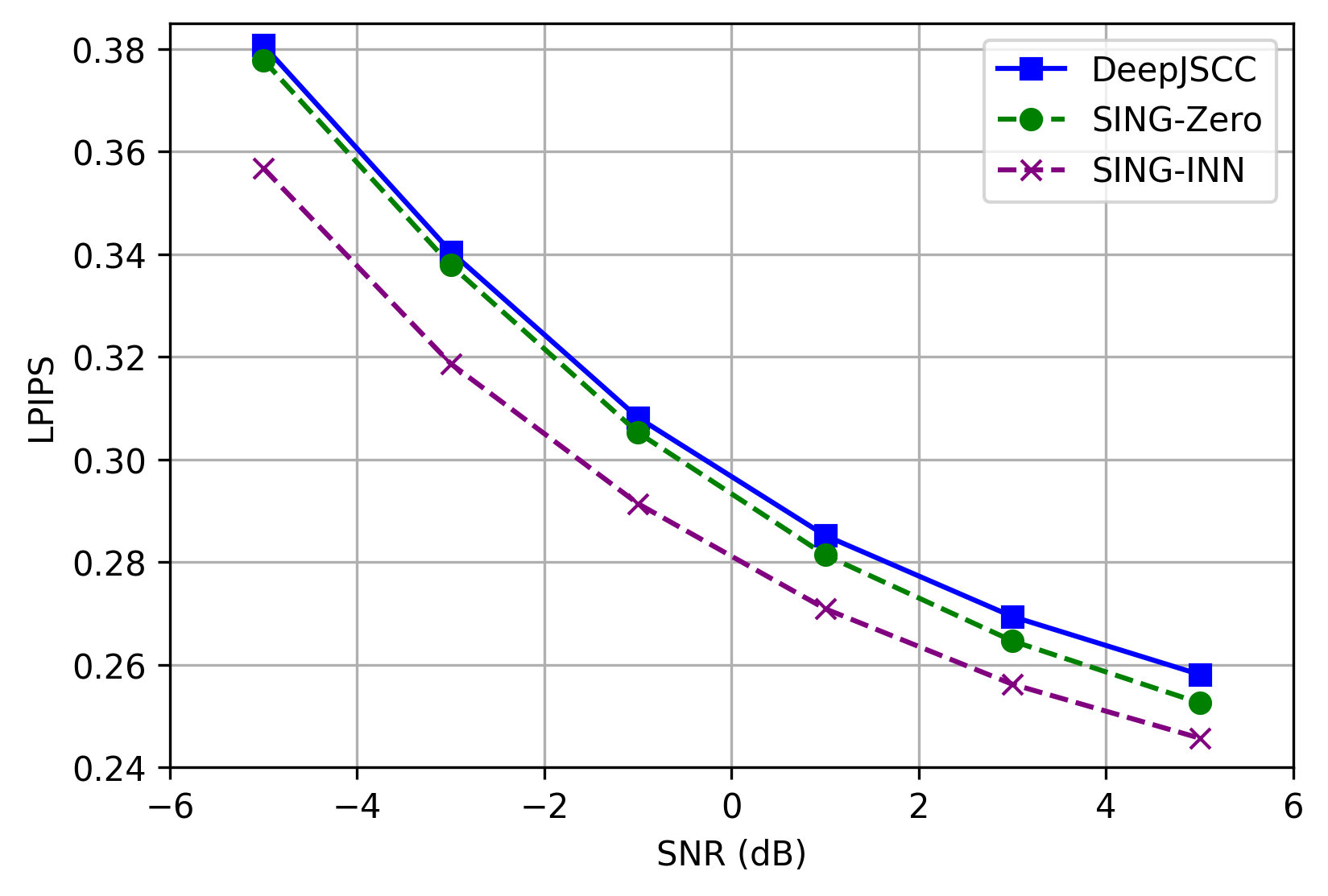}~\\
   (b) LPIPS versus SNR (lower better) \\
	\end{tabular}
    \vspace{-0.1cm} 
	\caption{Performance comparison in terms of PSNR and LPIPS for $\rho=0.0052$. } 
 \vspace{-0.1cm}
	\label{fig:performance0052} 
 \vspace{-0.3cm}
\end{figure}

\section{Experimental Results}
\label{sec:experimental_results}
To evaluate the performance of our proposed approach, we compare our two-stage method SING-Zero and SING-INN with the state-of-the-art JSCC methods: DeepJSCC~\cite{tung2021deepwive} and InverseJSCC~\cite{erdemir2022generative}. 
For fair comparisons, we use the same pre-trained DeepJSCC for all three approaches. We utilize the $512 \times 512$ CelebA-HQ dataset~\cite{karras2017progressive} following~\cite{erdemir2022generative}, comprising \num{30000} high-resolution celebrity images, divided into training, validation, and testing sets with a ratio of $8:1:1$.

\subsection{Implementation Details}
We apply standard hyperparameters for our method that are commonly used in the literature to train \gls{DeepJSCC}~\cite{bourtsoulatze2019deep,tung2022deepjscc}. Specifically, we use a learning rate of \num{1e-4}, a batch size of \num{32}, and an average power constraint of $\bar{P}=1.0$ (refer to \cref{eq:power_constraint}). For all the methods analyzed, PyTorch Adam optimizer~\cite{kingma2014adam} is utilized to minimize the training loss described in \cref{eq:ddpm_loss}. During training, the network is subjected to channel noise $\nv$ determined by a uniformly sampled SNR ranging from \num{-5} to \num{5} \si{\decibel}. Our INN is composed of 4 pairs of PNet/UNet, and each PNet/UNet consists of $J=2$ CondBlocks. For training INN, we use a learning rate of \num{5e-5} and a batch size of \num{32} on CelebA-HQ training dataset.

\begin{figure}[!t]\footnotesize %
\centering
\hspace{-0.26cm}
\begin{tabular}
{c@{\extracolsep{0em}}c@{\extracolsep{0em}}c}
       
		\includegraphics[width=0.35\textwidth]{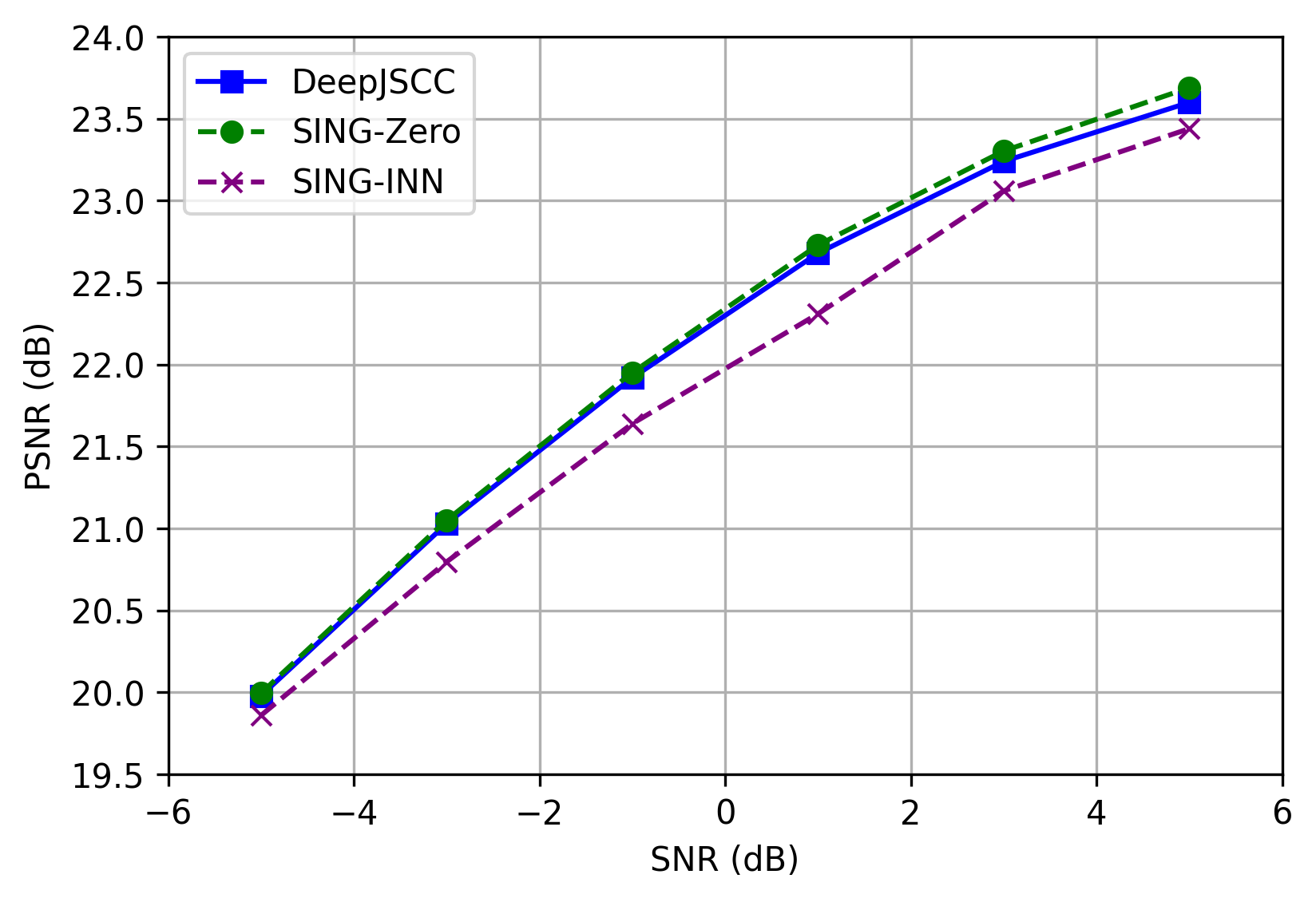}\\
     (a) PSNR versus SNR (higher better) \\ 
		\includegraphics[width=0.35\textwidth]{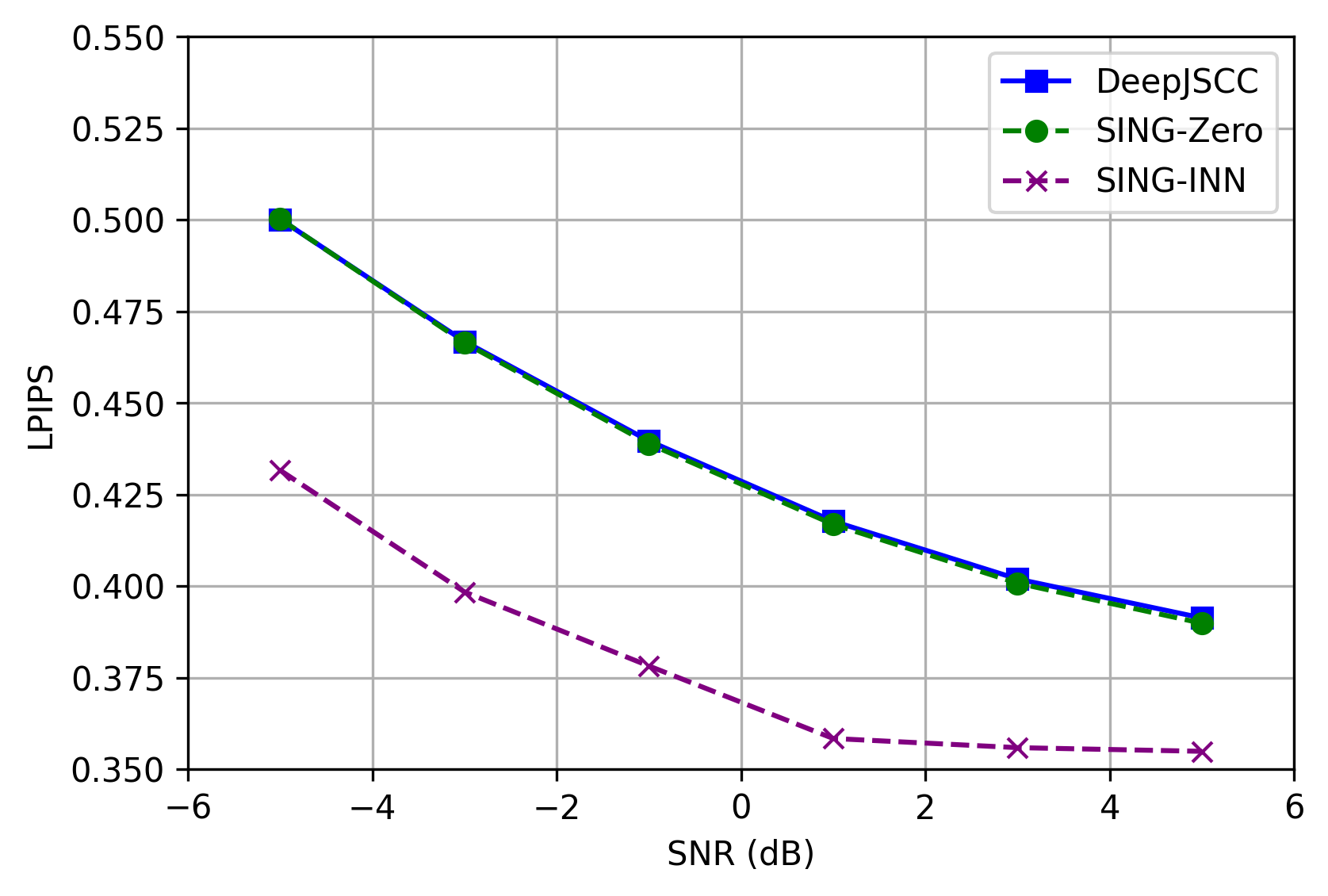}~\\
(b) LPIPS versus SNR (lower better) \\
	\end{tabular}
    \vspace{-0.1cm} 
	\caption{Performance comparison in terms of PSNR and LPIPS for $\rho=0.0013$. } 
 \vspace{-0.1cm}
	\label{fig:performance0013} 
 \vspace{-0.3cm}
\end{figure}

We assume perfect channel knowledge at the encoder and the decoder. We experiment on channel SNRs of $\{-5,-3,-1,1,3,5\}$ dB and BCR values $\rho=\{ 0.0013, 0.0052 \}$, which correspond to highly challenging communication scenarios. 
Empirically, we set $T=1000$ and step size $\zeta$ to $0.3$, $0.4$, and $0.5$
for SNR = $\{-5, -3\}$ dB, $\{-1, 1\}$ dB, and $\{3, 5\}$ dB respectively.
We analyze the performance of the proposed scheme using both the widely used pixel-wise metric PSNR and the perceptual metric LPIPS~\cite{zhang2018unreasonable}.

\begin{figure*}[p]\footnotesize %
\begin{center}
    \begin{tabular}{@{}c@{}c@{}c@{}c@{}c@{}c@{}}
        \includegraphics[width=0.16\textwidth]{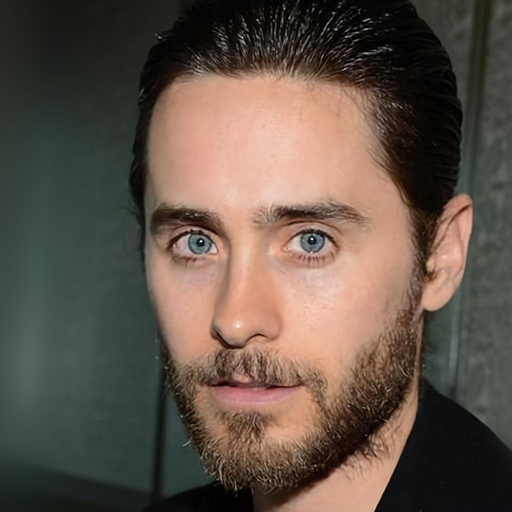} & 
        \hspace{0.001\textwidth}
        \includegraphics[width=0.16\textwidth]{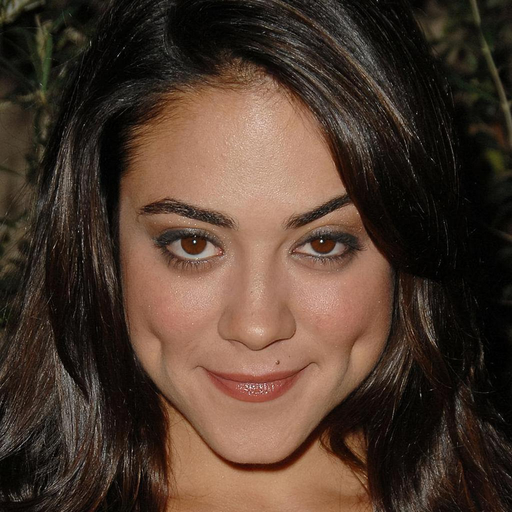} & 
        \hspace{0.001\textwidth}
        \includegraphics[width=0.16\textwidth]{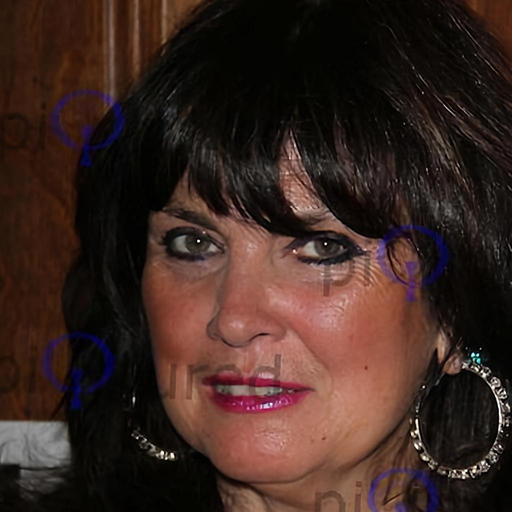} & 
        \hspace{0.001\textwidth}
        \includegraphics[width=0.16\textwidth]{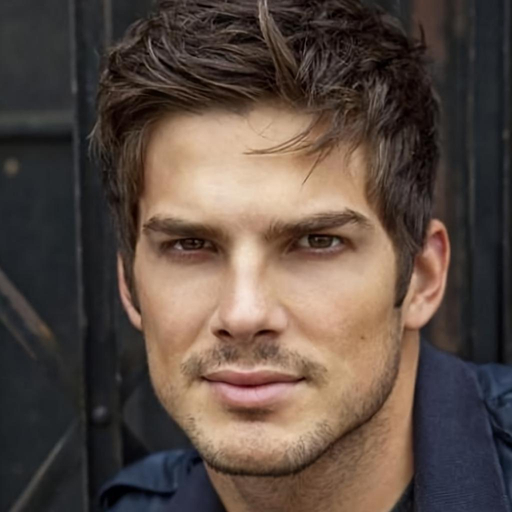} & 
        \hspace{0.001\textwidth}
        \includegraphics[width=0.16\textwidth]{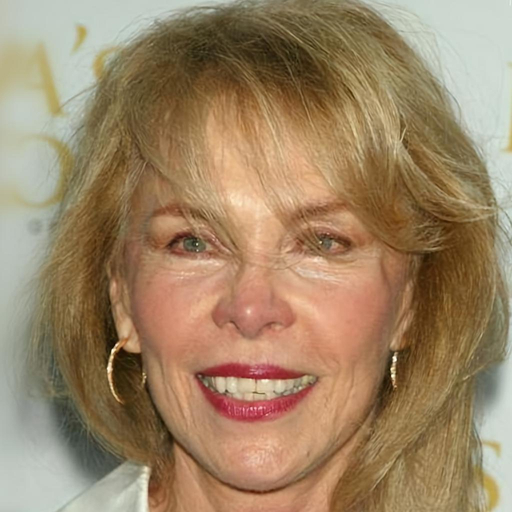} & 
        \hspace{0.001\textwidth}
        \includegraphics[width=0.16\textwidth]{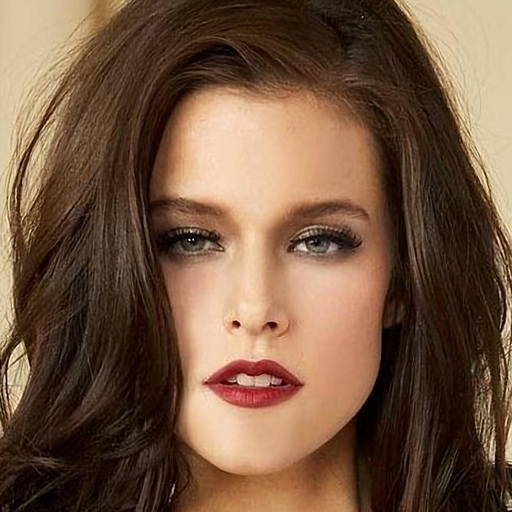}  \\
        \multicolumn{6}{c}{(a) Ground Truth.} \\
                        \includegraphics[width=0.16\textwidth]{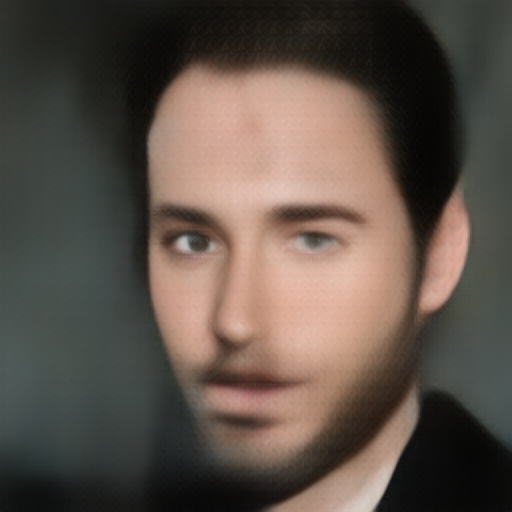} & 
        \hspace{0.001\textwidth}
        \includegraphics[width=0.16\textwidth]{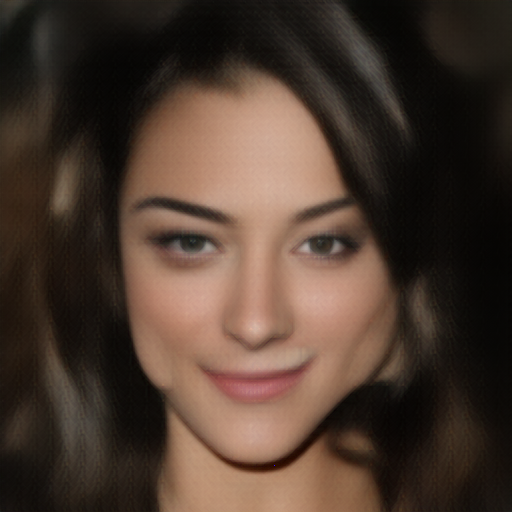} & 
        \hspace{0.001\textwidth}
        \includegraphics[width=0.16\textwidth]{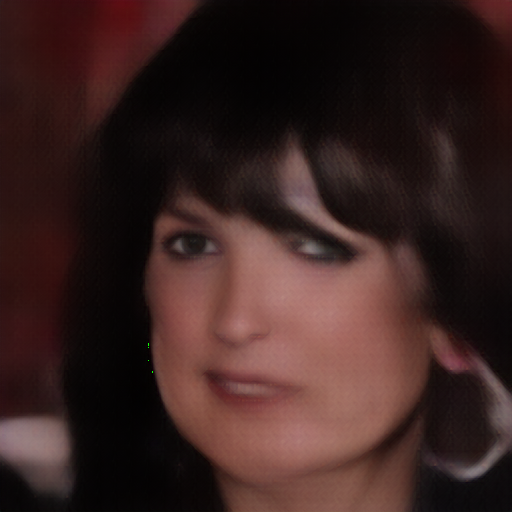} & 
        \hspace{0.001\textwidth}
        \includegraphics[width=0.16\textwidth]{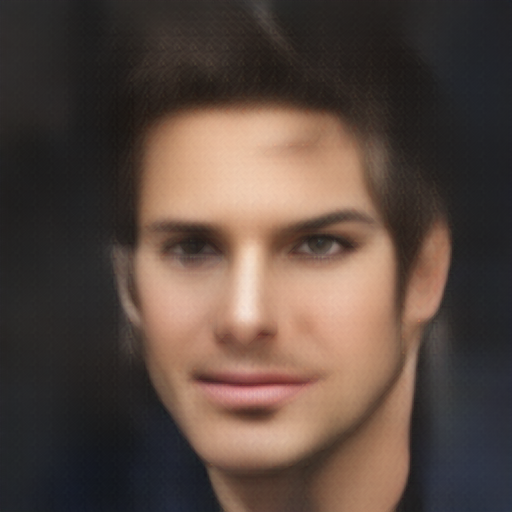} & 
        \hspace{0.001\textwidth}
        \includegraphics[width=0.16\textwidth]{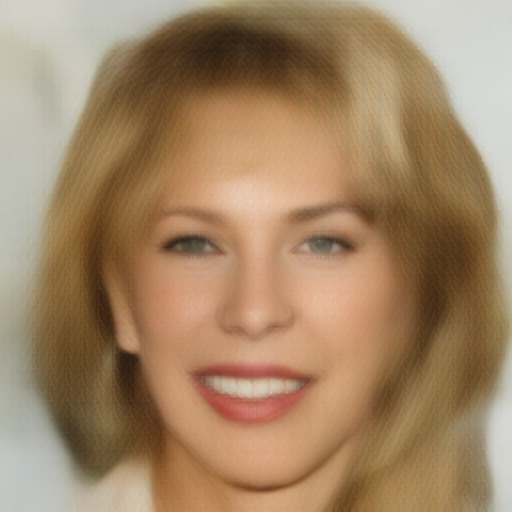} & 
        \hspace{0.001\textwidth}
        \includegraphics[width=0.16\textwidth]{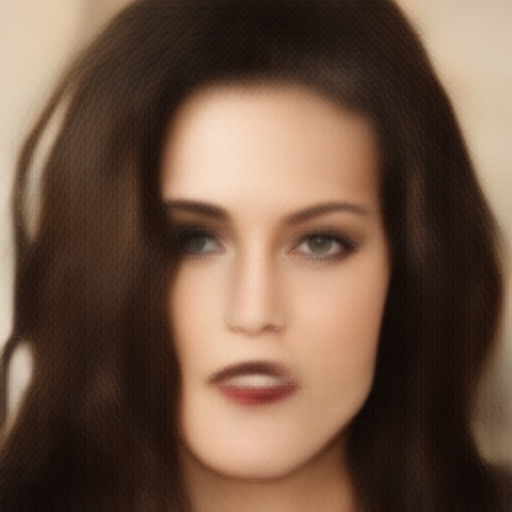}  \\
        \multicolumn{6}{c}{(b) DeepJSCC at SNR=$-1$~dB.} \\

                        \includegraphics[width=0.16\textwidth]{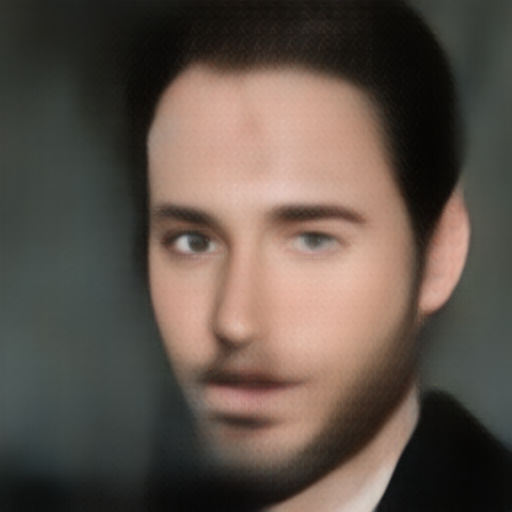} & 
        \hspace{0.001\textwidth}
        \includegraphics[width=0.16\textwidth]{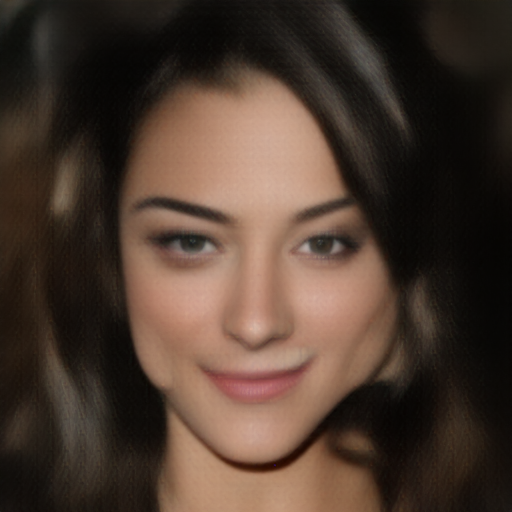} & 
        \hspace{0.001\textwidth}
        \includegraphics[width=0.16\textwidth]{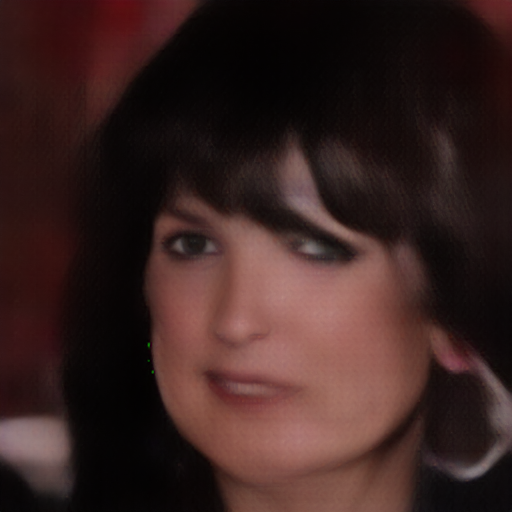} & 
        \hspace{0.001\textwidth}
        \includegraphics[width=0.16\textwidth]{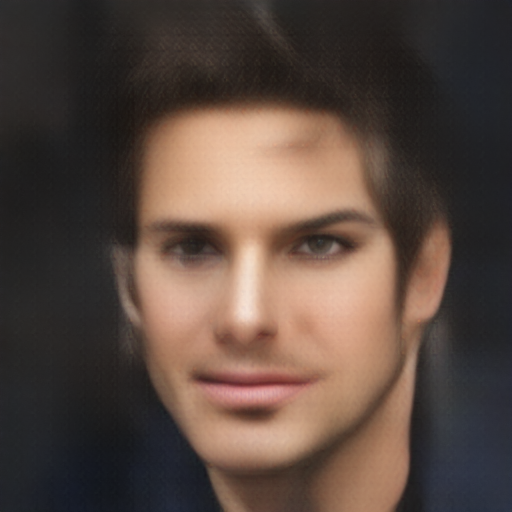} & 
        \hspace{0.001\textwidth}
        \includegraphics[width=0.16\textwidth]{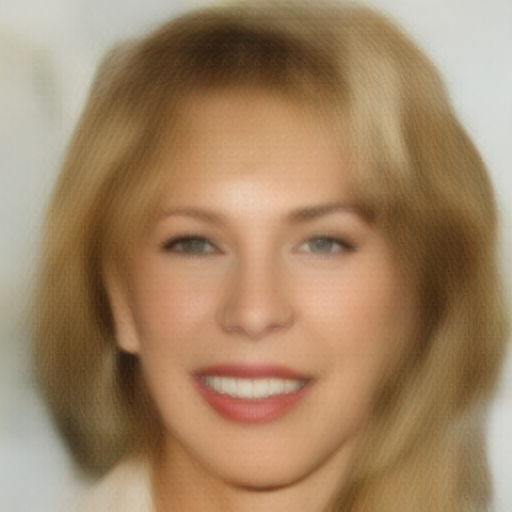} & 
        \hspace{0.001\textwidth}
        \includegraphics[width=0.16\textwidth]{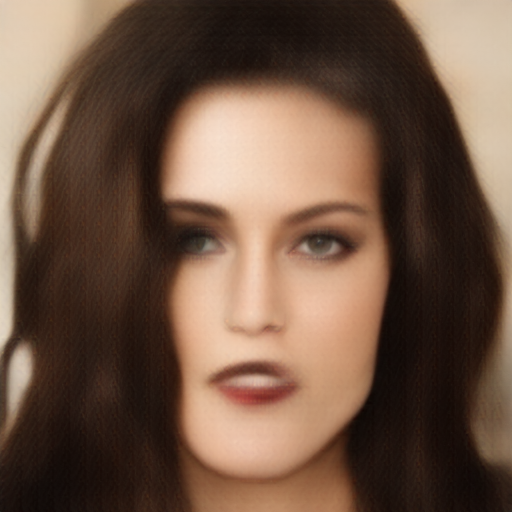}  \\
        \multicolumn{6}{c}{(c) SING-Zero at SNR=$-1$~dB.} \\

        \includegraphics[width=0.16\textwidth]{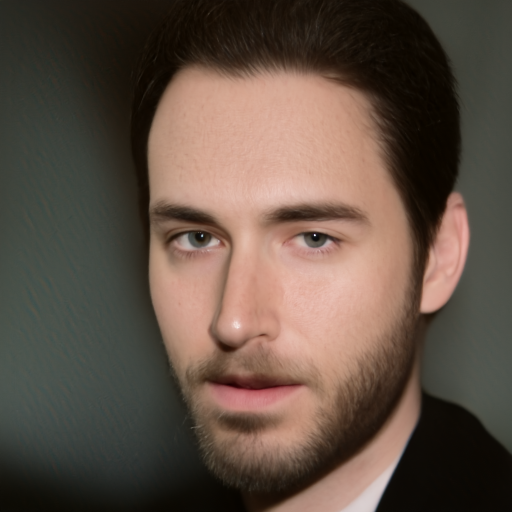} & 
        \hspace{0.001\textwidth}
        \includegraphics[width=0.16\textwidth]{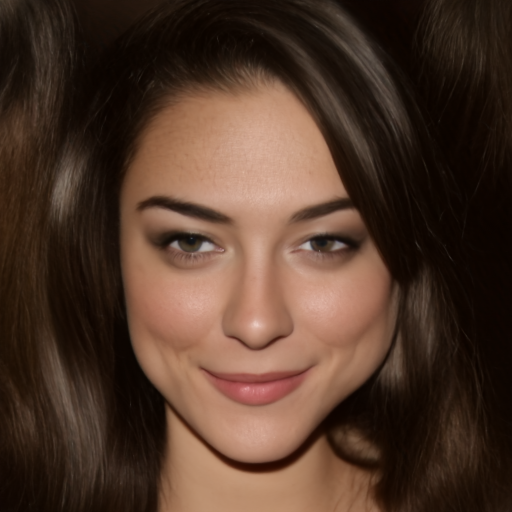} & 
        \hspace{0.001\textwidth}
        \includegraphics[width=0.16\textwidth]{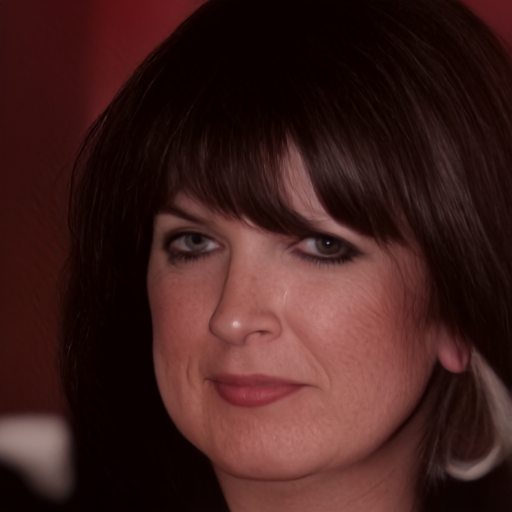} & 
        \hspace{0.001\textwidth}
        \includegraphics[width=0.16\textwidth]{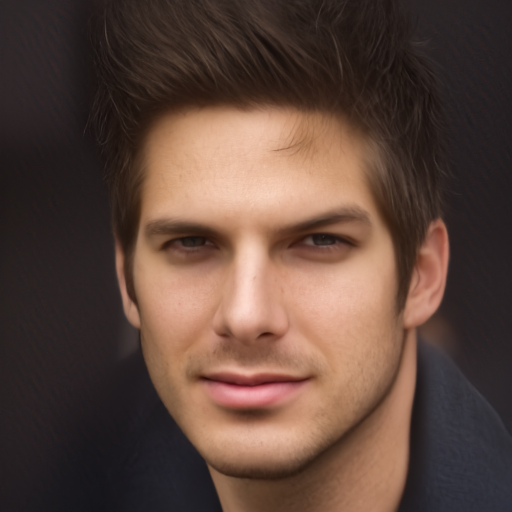} & 
        \hspace{0.001\textwidth}
        \includegraphics[width=0.16\textwidth]{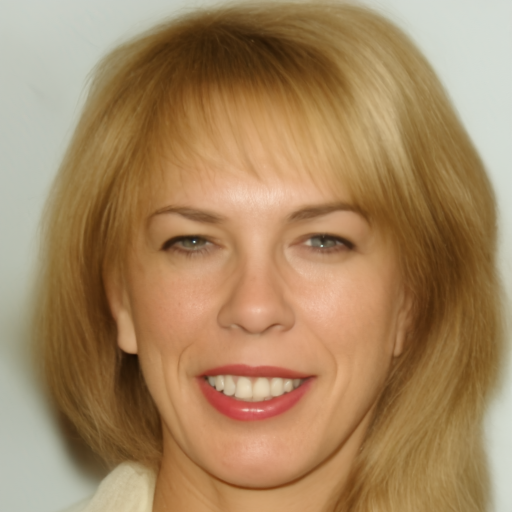} & 
        \hspace{0.001\textwidth}
        \includegraphics[width=0.16\textwidth]{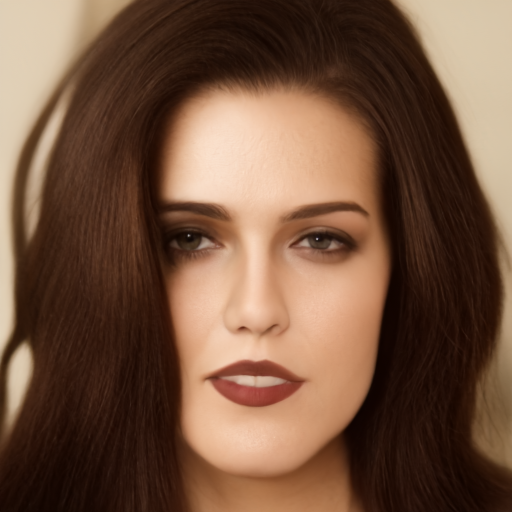}  \\
        \multicolumn{6}{c}{(d) SING-INN at SNR=$-1$~dB.} \\
                \includegraphics[width=0.16\textwidth]{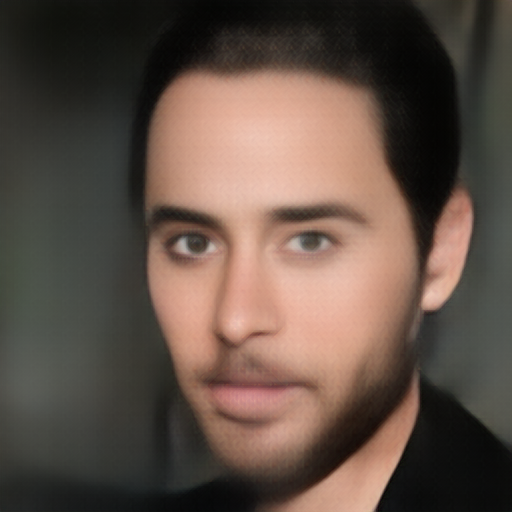} & 
        \hspace{0.001\textwidth}
        \includegraphics[width=0.16\textwidth]{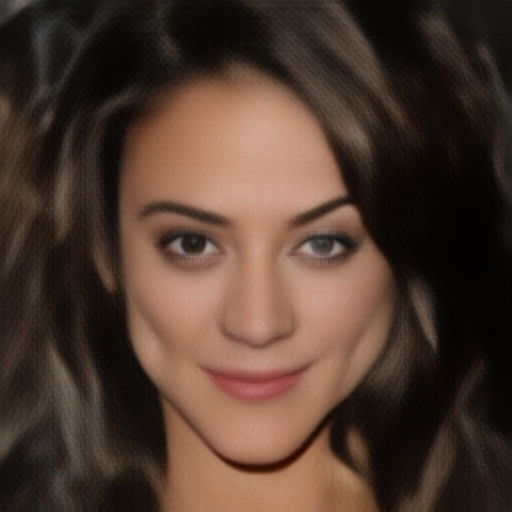} & 
        \hspace{0.001\textwidth}
        \includegraphics[width=0.16\textwidth]{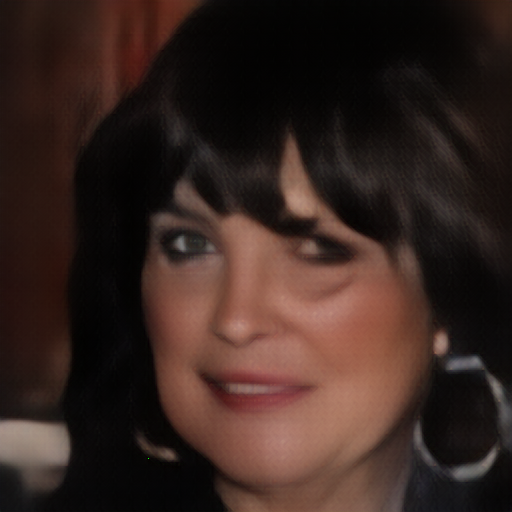} & 
        \hspace{0.001\textwidth}
        \includegraphics[width=0.16\textwidth]{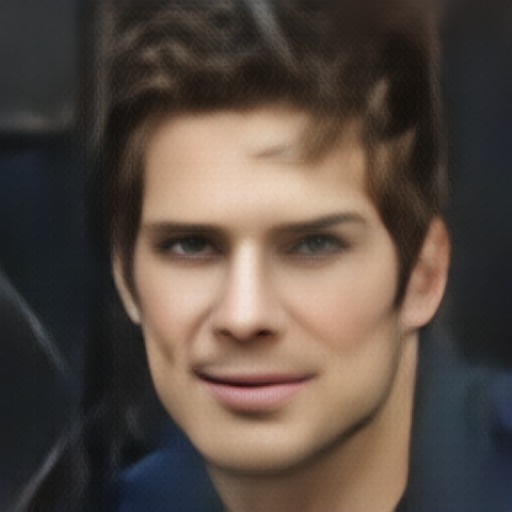} & 
        \hspace{0.001\textwidth}
        \includegraphics[width=0.16\textwidth]{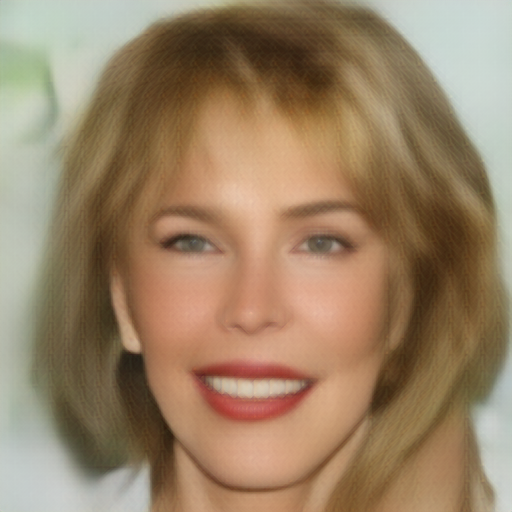} & 
        \hspace{0.001\textwidth}
        \includegraphics[width=0.16\textwidth]{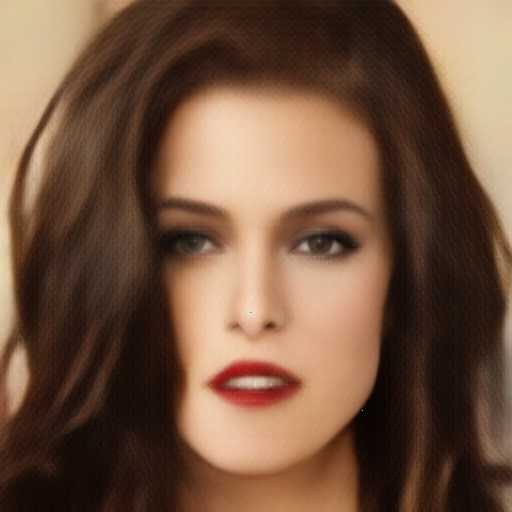}  \\
        \multicolumn{6}{c}{(e) DeepJSCC at SNR=$5$~dB.} \\

                \includegraphics[width=0.16\textwidth]{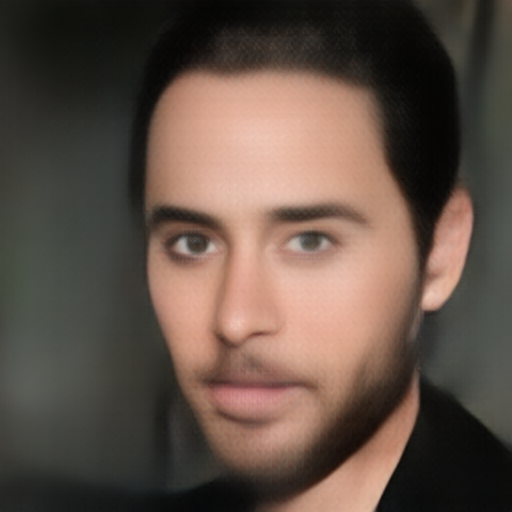} & 
        \hspace{0.001\textwidth}
        \includegraphics[width=0.16\textwidth]{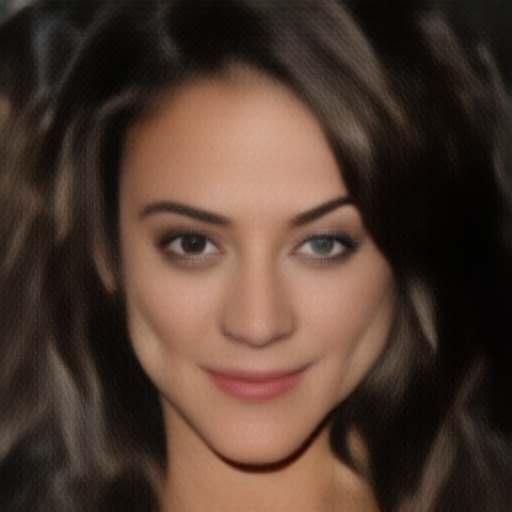} & 
        \hspace{0.001\textwidth}
        \includegraphics[width=0.16\textwidth]{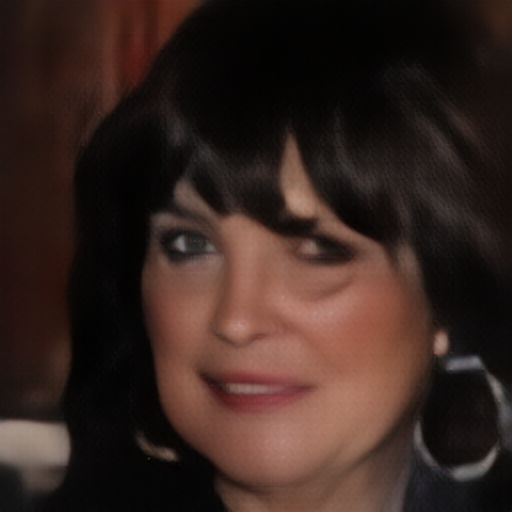} & 
        \hspace{0.001\textwidth}
        \includegraphics[width=0.16\textwidth]{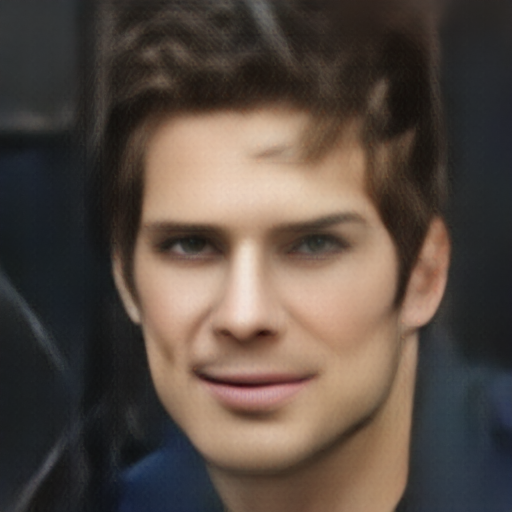} & 
        \hspace{0.001\textwidth}
        \includegraphics[width=0.16\textwidth]{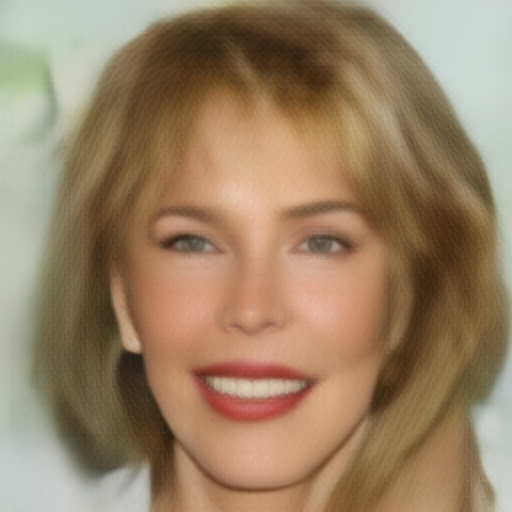} & 
        \hspace{0.001\textwidth}
        \includegraphics[width=0.16\textwidth]{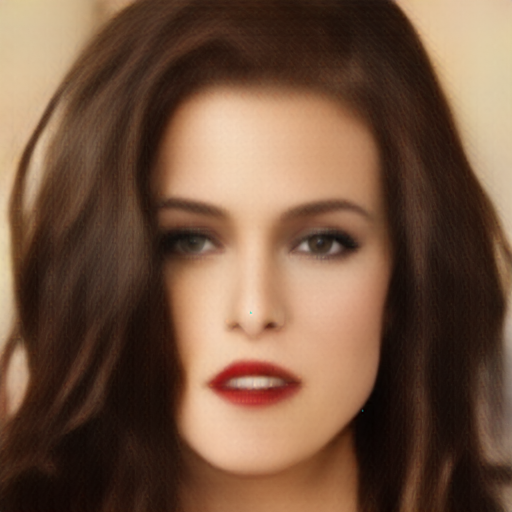}  \\
        \multicolumn{6}{c}{(f) SING-Zero at SNR=$5$~dB.} \\

        \includegraphics[width=0.16\textwidth]{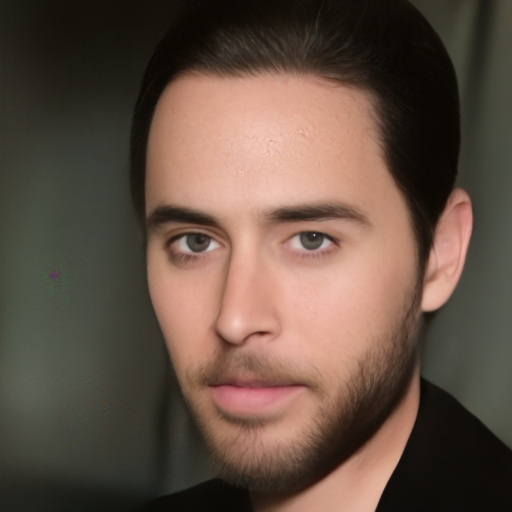} & 
        \hspace{0.001\textwidth}
        \includegraphics[width=0.16\textwidth]{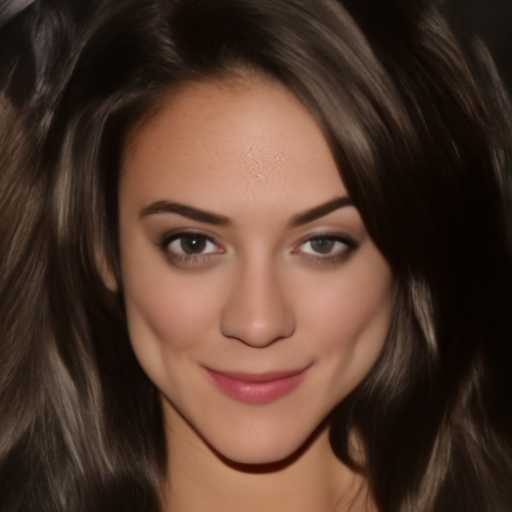} & 
        \hspace{0.001\textwidth}
        \includegraphics[width=0.16\textwidth]{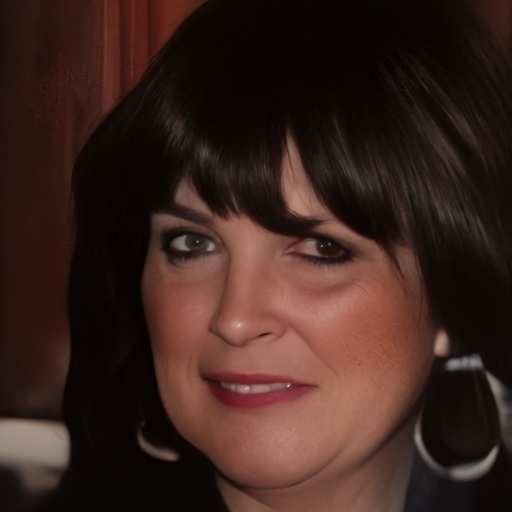} & 
        \hspace{0.001\textwidth}
        \includegraphics[width=0.16\textwidth]{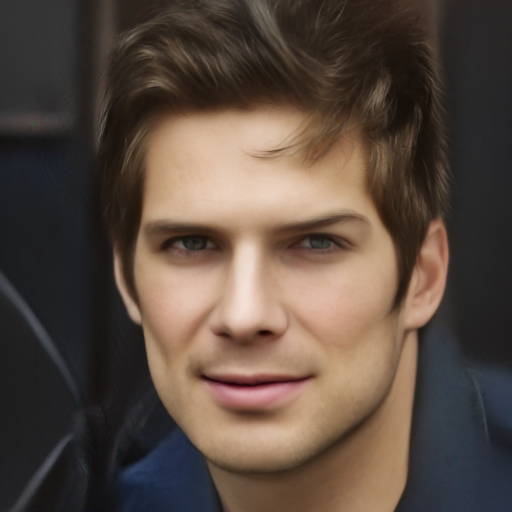} & 
        \hspace{0.001\textwidth}
        \includegraphics[width=0.16\textwidth]{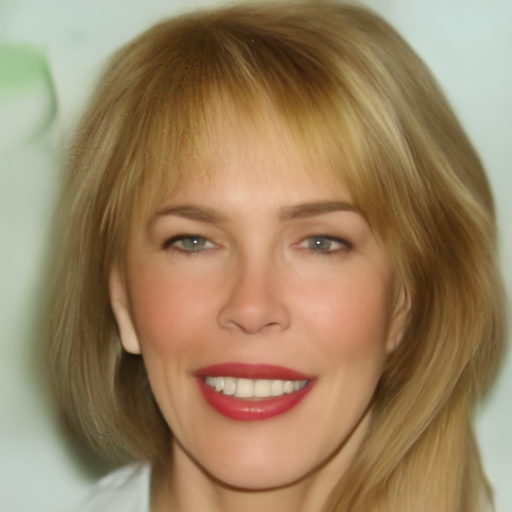} & 
        \hspace{0.001\textwidth}
        \includegraphics[width=0.16\textwidth]{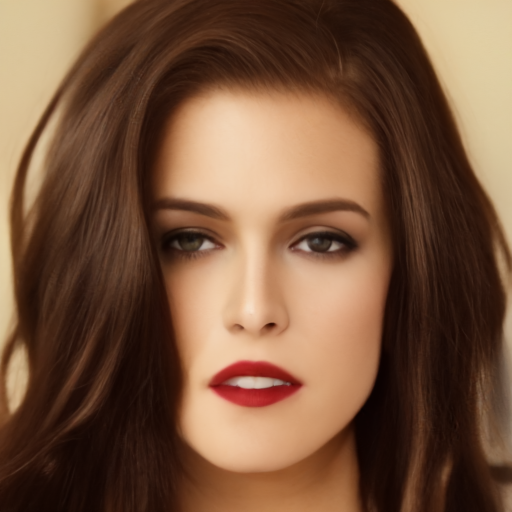}  \\
        \multicolumn{6}{c}{(g) SING-INN at SNR=$5$~dB.} \\
    \end{tabular}
   
\end{center}
\caption{Original and reconstructed CelebA-HQ images by DeepJSCC and our proposed SING-Zero and SING-INN for $\rho=0.0013$ and SNR=$\{-1, 5\}$. }
\label{fig:Visual Results}
\end{figure*}
\clearpage 

\begin{figure*}[p]\footnotesize %
\begin{center}
    \begin{tabular}{@{}c@{}c@{}c@{}c@{}c@{}}
        \includegraphics[width=0.19\textwidth]{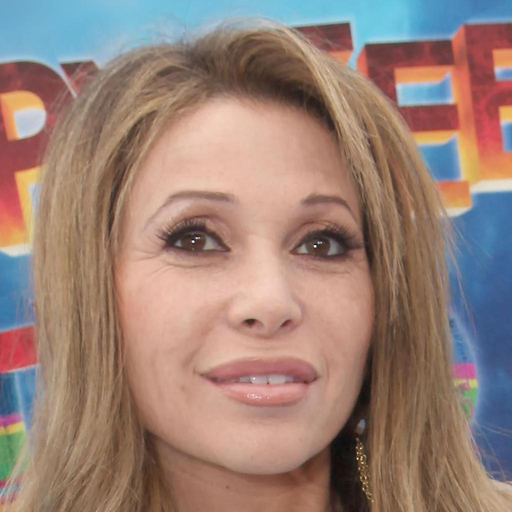} & 
        \includegraphics[width=0.19\textwidth]{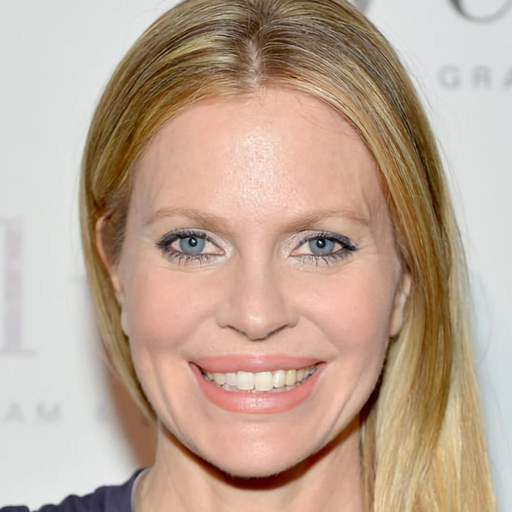} & 
        \includegraphics[width=0.19\textwidth]{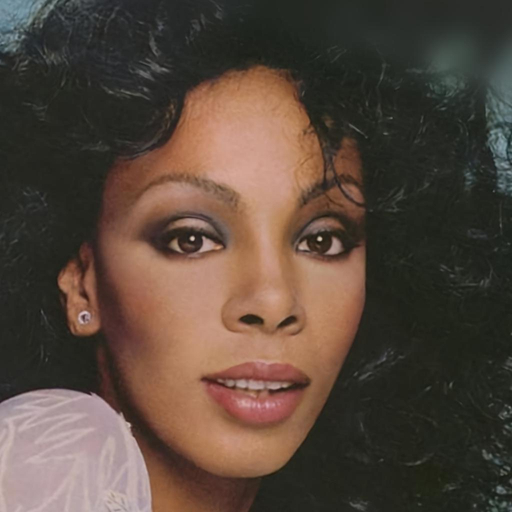} & 
        \includegraphics[width=0.19\textwidth]{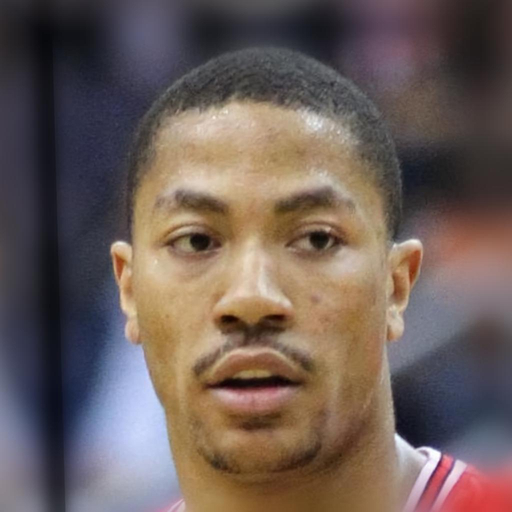} & 
        \includegraphics[width=0.19\textwidth]{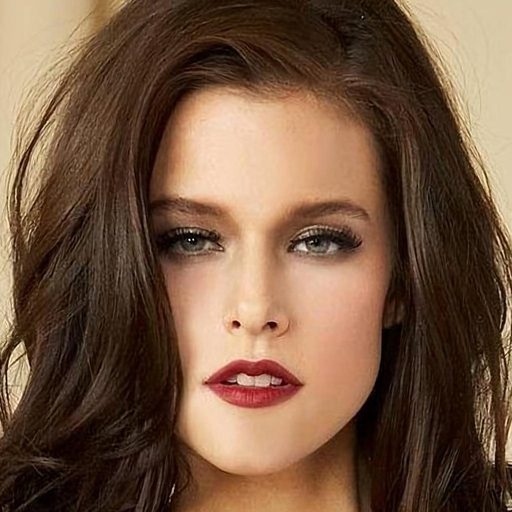} \\
        \multicolumn{5}{c}{(a) Ground Truth.} \\
        \includegraphics[width=0.19\textwidth]{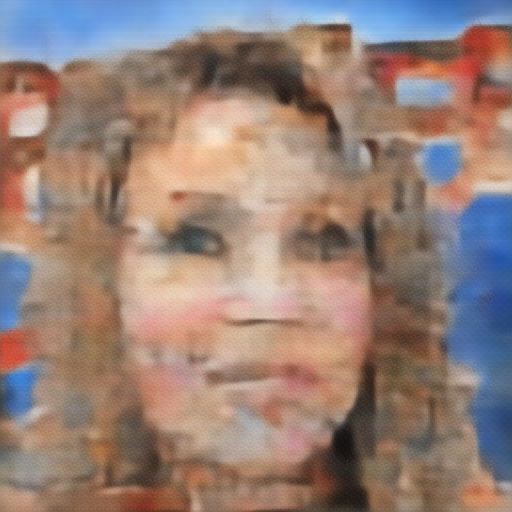} & 
        \includegraphics[width=0.19\textwidth]{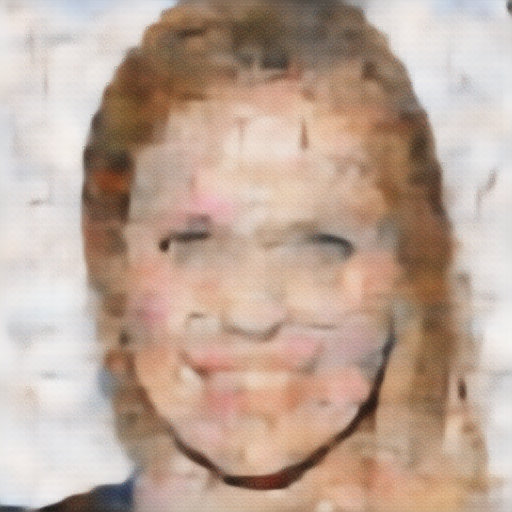} & 
        \includegraphics[width=0.19\textwidth]{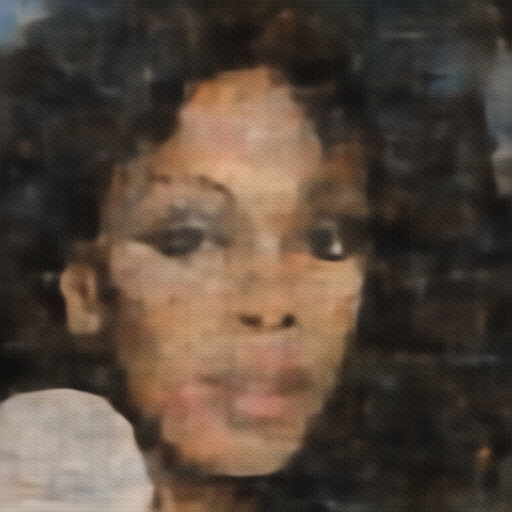} & 
        \includegraphics[width=0.19\textwidth]{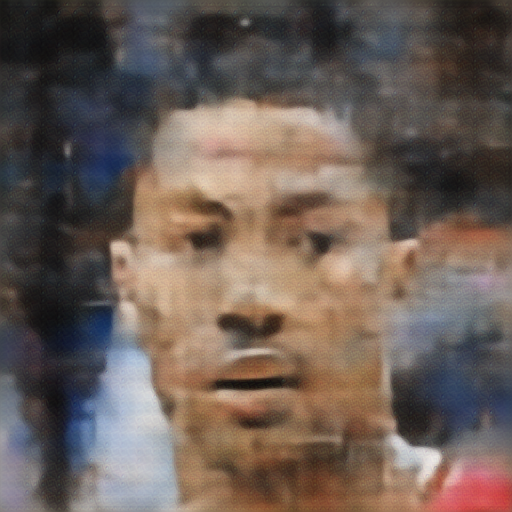} & 
        \includegraphics[width=0.19\textwidth]{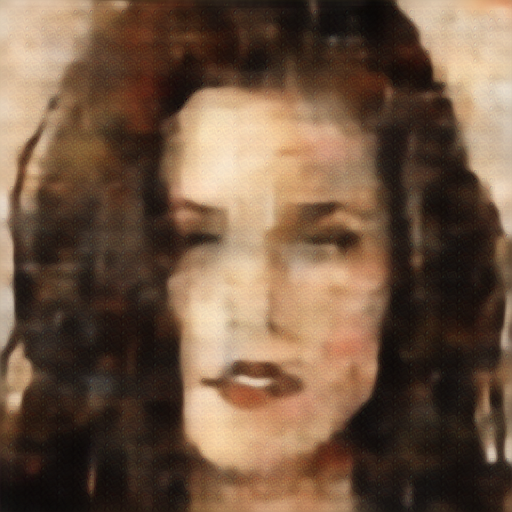} \\
        \multicolumn{5}{c}{(b) DeepJSCC at SNR=$1$~dB.} \\
        \includegraphics[width=0.19\textwidth]{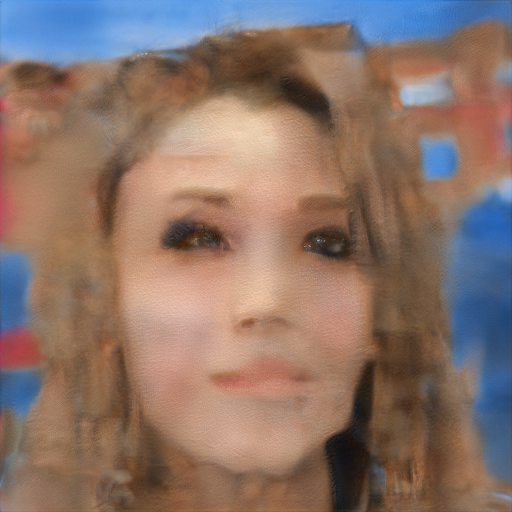} & 
        \includegraphics[width=0.19\textwidth]{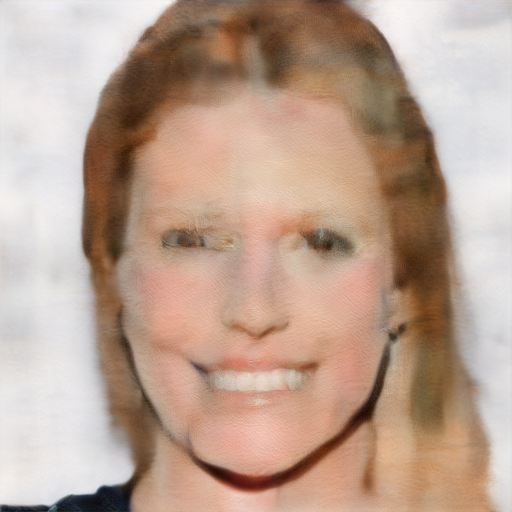} & 
        \includegraphics[width=0.19\textwidth]{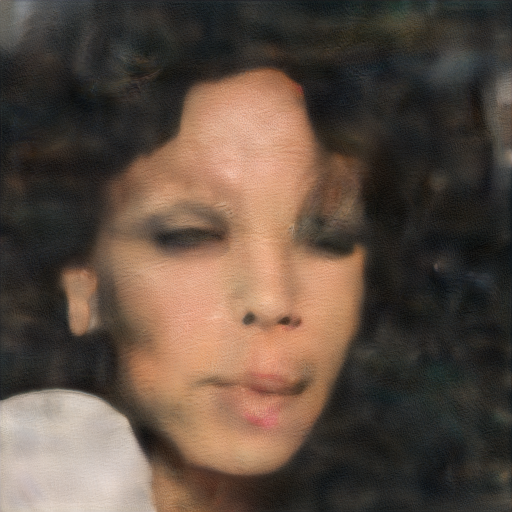} & 
        \includegraphics[width=0.19\textwidth]{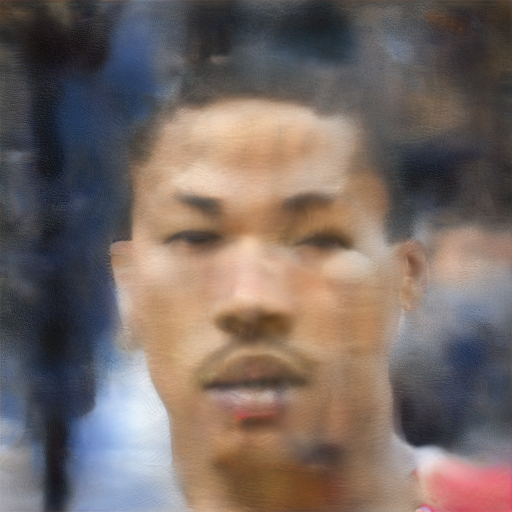} & 
        \includegraphics[width=0.19\textwidth]{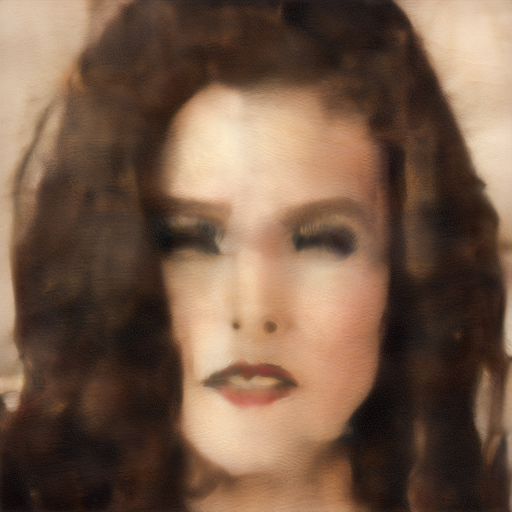} \\
        \multicolumn{5}{c}{(c) InverseJSCC at SNR=$1$~dB.} \\
        \includegraphics[width=0.19\textwidth]{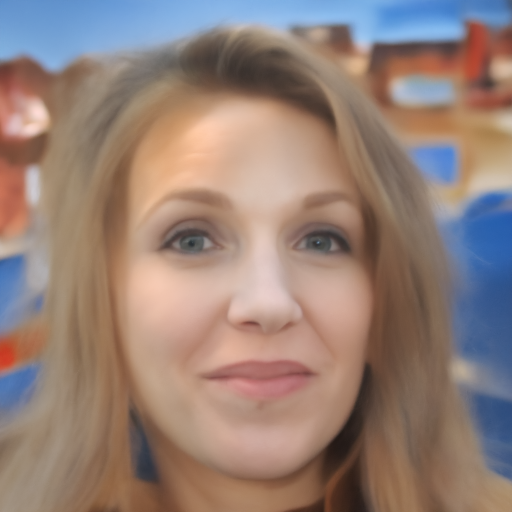} & 
        \includegraphics[width=0.19\textwidth]{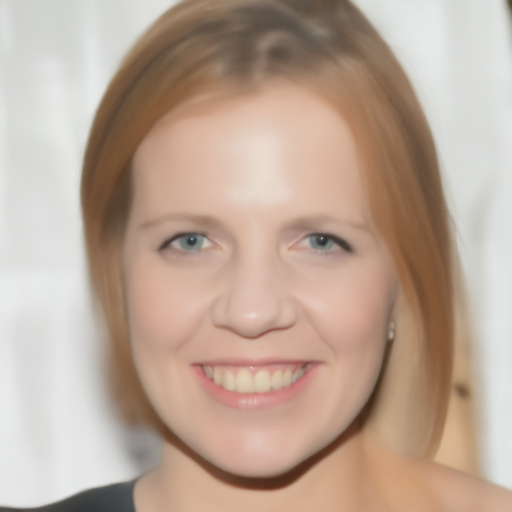} & 
        \includegraphics[width=0.19\textwidth]{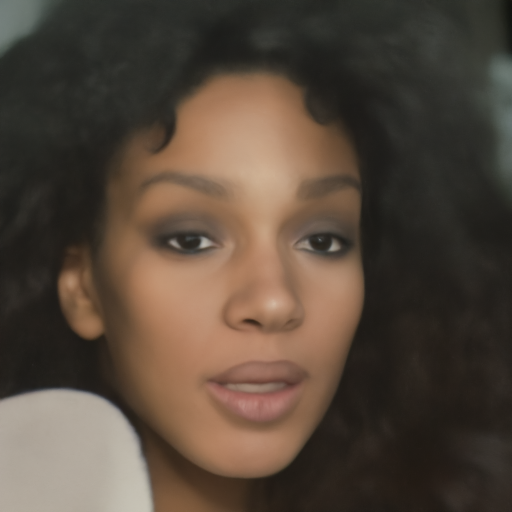} & 
        \includegraphics[width=0.19\textwidth]{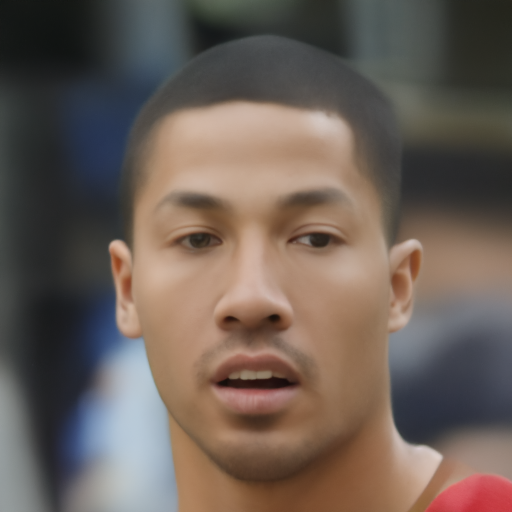} & 
        \includegraphics[width=0.19\textwidth]{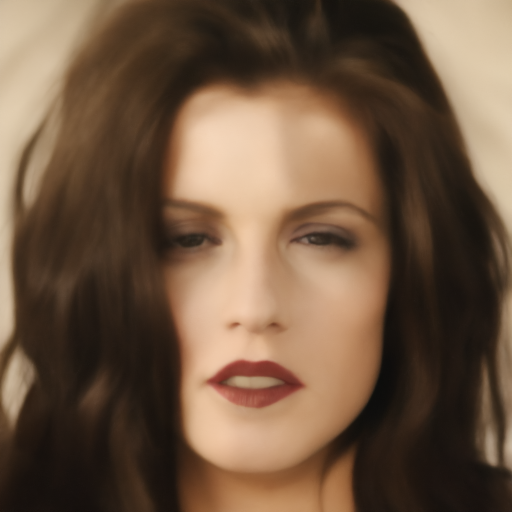} \\
        \multicolumn{5}{c}{(d) SING-Zero at SNR=$1$~dB.} \\
        \includegraphics[width=0.19\textwidth]{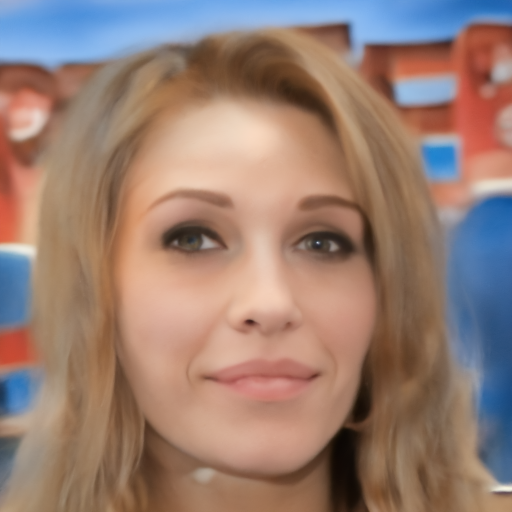} & 
        \includegraphics[width=0.19\textwidth]{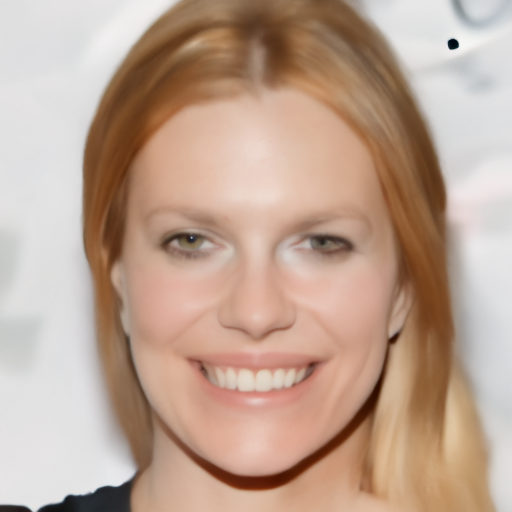} & 
        \includegraphics[width=0.19\textwidth]{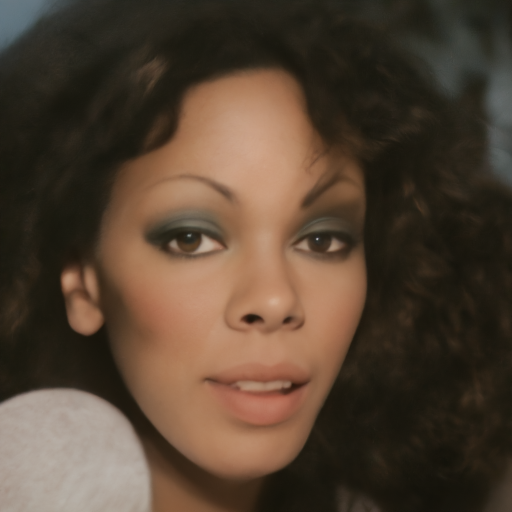} & 
        \includegraphics[width=0.19\textwidth]{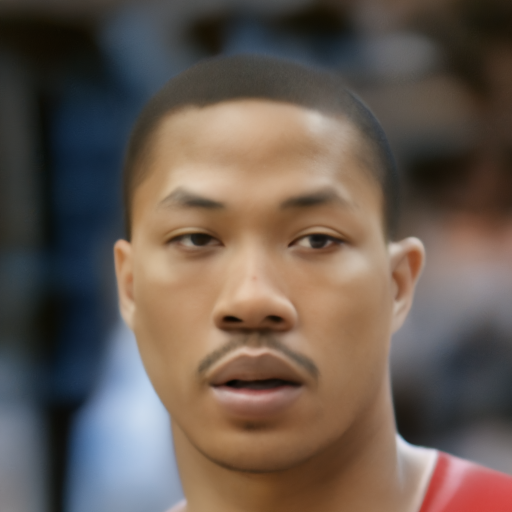} & 
        \includegraphics[width=0.19\textwidth]{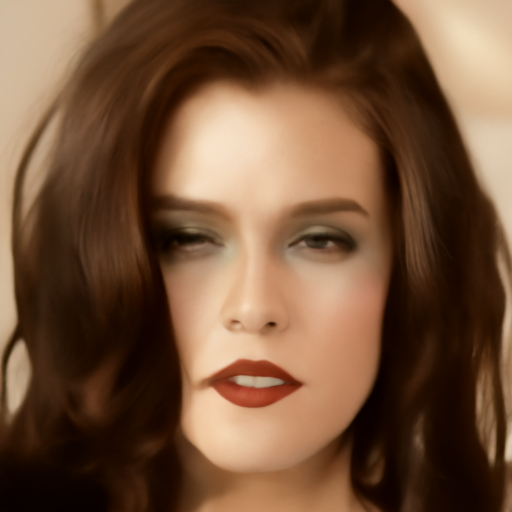} \\
        \multicolumn{5}{c}{(e) SING-INN at SNR=$1$~dB.} \\

    \end{tabular}
\end{center}
\caption{Original and reconstructed CelebA-HQ images using DeepJSCC trained on TinyImageNet, along with results from InverseJSCC, SING-Zero, and SING-INN under $\rho=0.0052$ and SNR=$1$~dB.}
\label{fig:Visual Results Tiny1}
\end{figure*}
\clearpage
\begin{figure*}[p]\footnotesize %
\begin{center}
    \begin{tabular}{@{}c@{}c@{}c@{}c@{}c@{}}
        \includegraphics[width=0.19\textwidth]{figure/visual_res_tiny/gt/1.png} & 
        \includegraphics[width=0.19\textwidth]{figure/visual_res_tiny/gt/2.png} & 
        \includegraphics[width=0.19\textwidth]{figure/visual_res_tiny/gt/3.png} & 
        \includegraphics[width=0.19\textwidth]{figure/visual_res_tiny/gt/4.png} & 
        \includegraphics[width=0.19\textwidth]{figure/visual_res_tiny/gt/5.png} \\
        \multicolumn{5}{c}{(a) Ground Truth.} \\
        \includegraphics[width=0.19\textwidth]{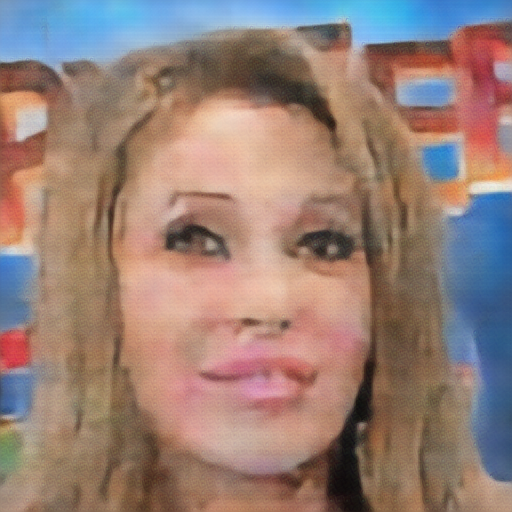} & 
        \includegraphics[width=0.19\textwidth]{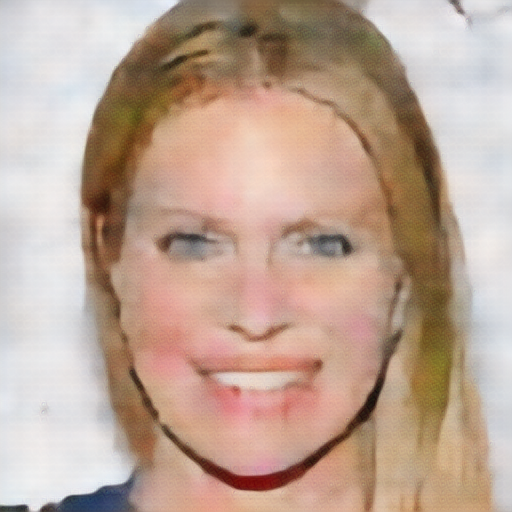} & 
        \includegraphics[width=0.19\textwidth]{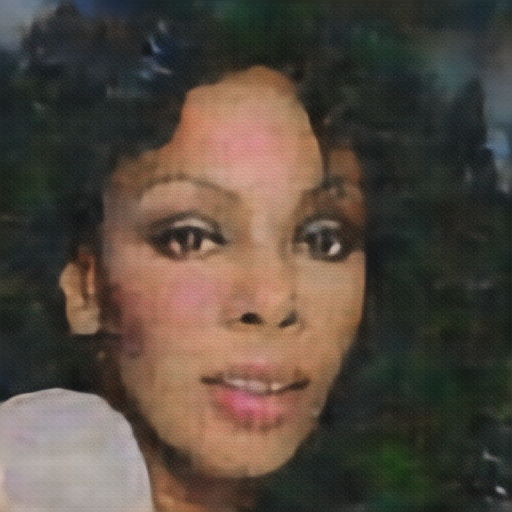} & 
        \includegraphics[width=0.19\textwidth]{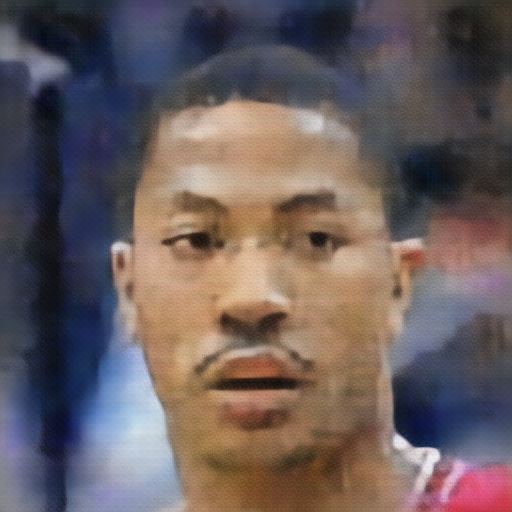} & 
        \includegraphics[width=0.19\textwidth]{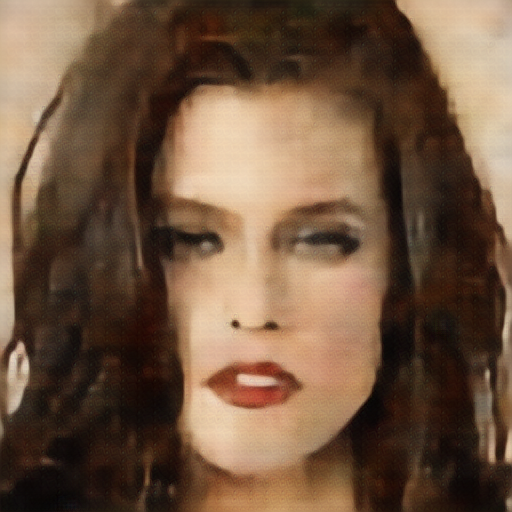} \\
        \multicolumn{5}{c}{(b) DeepJSCC at SNR=$5$~dB.} \\
        \includegraphics[width=0.19\textwidth]{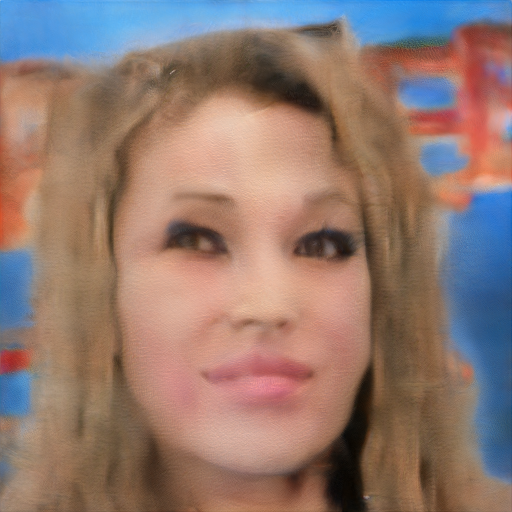} & 
        \includegraphics[width=0.19\textwidth]{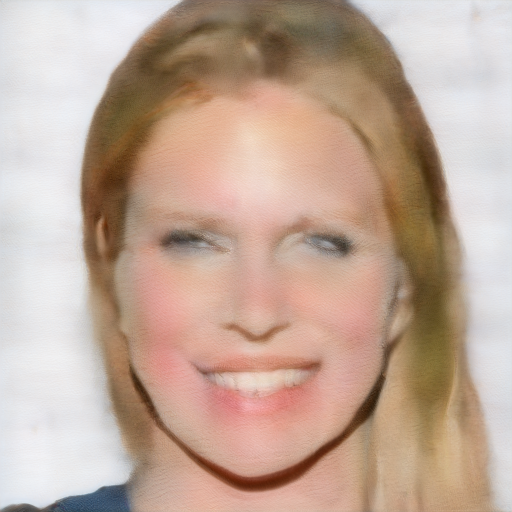} & 
        \includegraphics[width=0.19\textwidth]{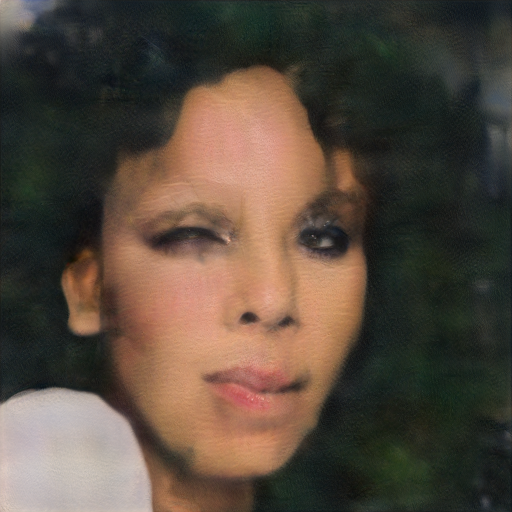} & 
        \includegraphics[width=0.19\textwidth]{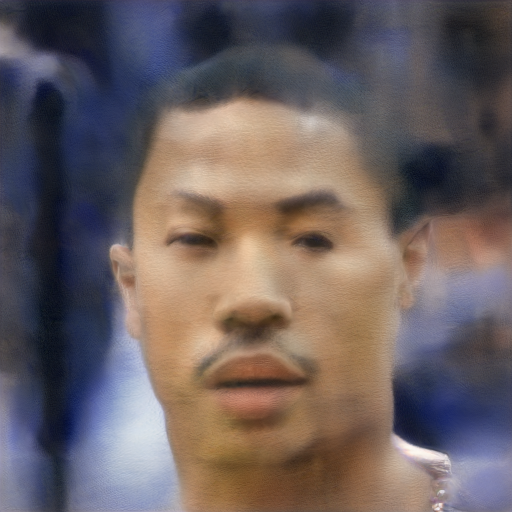} & 
        \includegraphics[width=0.19\textwidth]{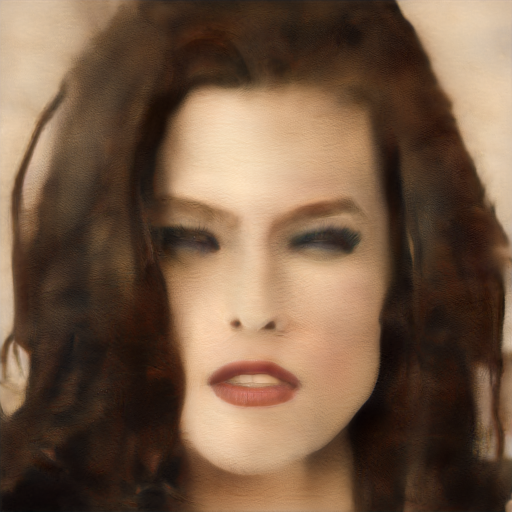} \\
        \multicolumn{5}{c}{(c) InverseJSCC at SNR=$5$~dB.} \\
        \includegraphics[width=0.19\textwidth]{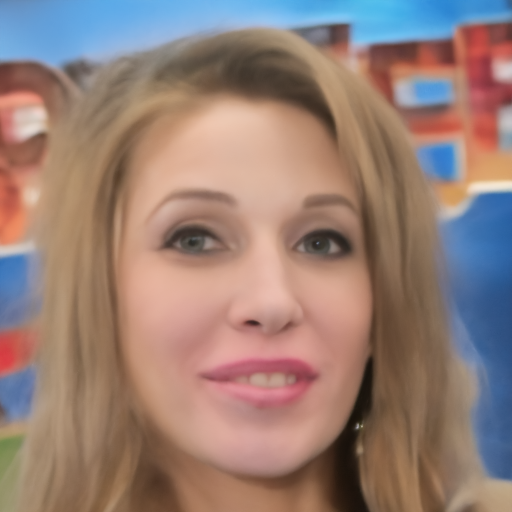} & 
        \includegraphics[width=0.19\textwidth]{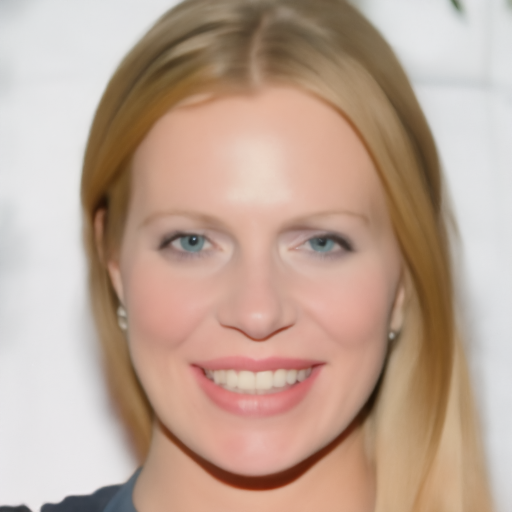} & 
        \includegraphics[width=0.19\textwidth]{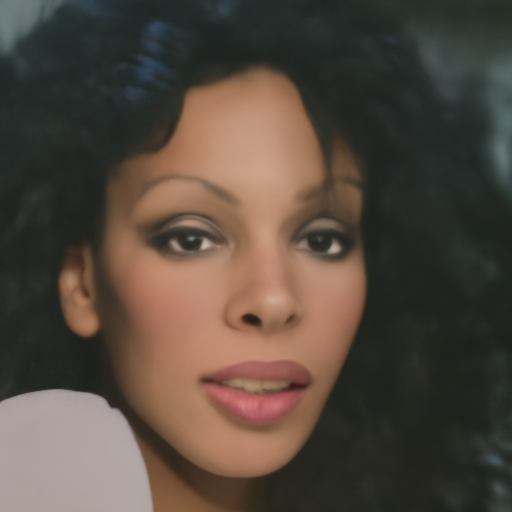} & 
        \includegraphics[width=0.19\textwidth]{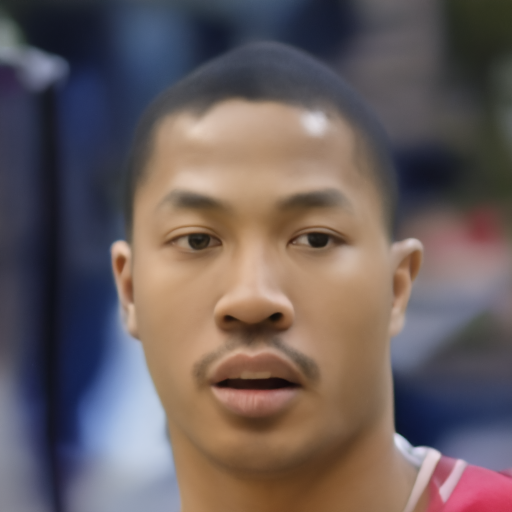} & 
        \includegraphics[width=0.19\textwidth]{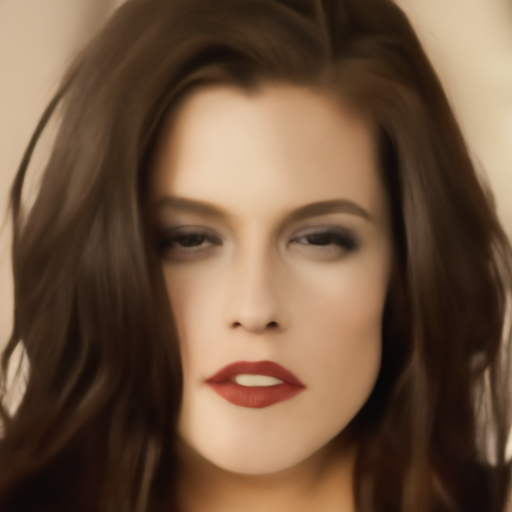} \\
        \multicolumn{5}{c}{(d) SING-Zero at SNR=$5$~dB.} \\
        \includegraphics[width=0.19\textwidth]{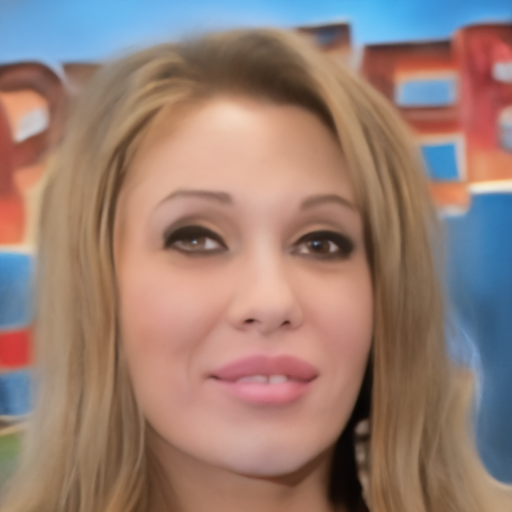} & 
        \includegraphics[width=0.19\textwidth]{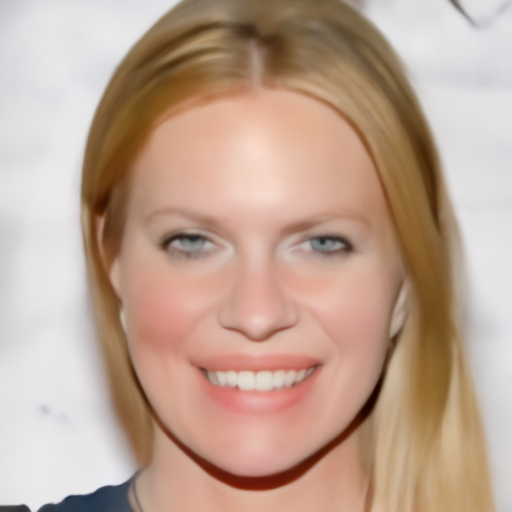} & 
        \includegraphics[width=0.19\textwidth]{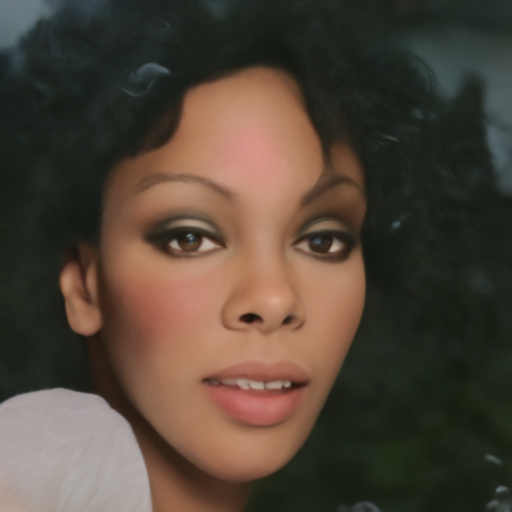} & 
        \includegraphics[width=0.19\textwidth]{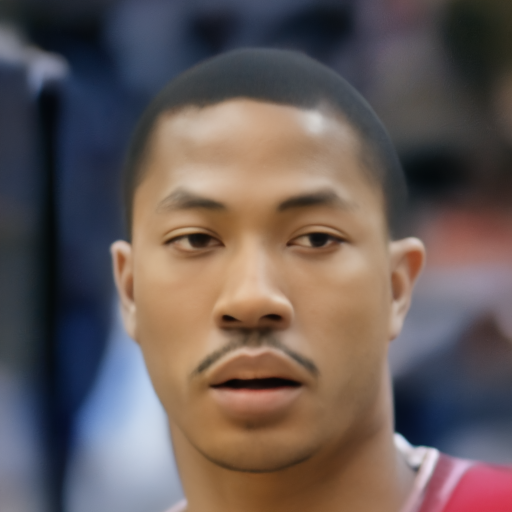} & 
        \includegraphics[width=0.19\textwidth]{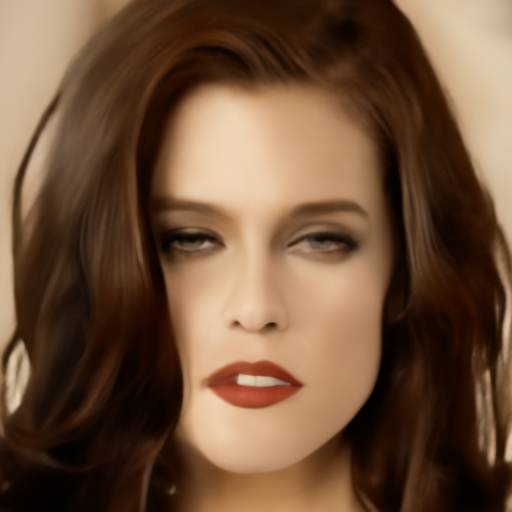} \\
        \multicolumn{5}{c}{(e) SING-INN at SNR=$5$~dB.} \\

    \end{tabular}
\end{center}
\caption{Original and reconstructed CelebA-HQ images using DeepJSCC trained on TinyImageNet, along with results from InverseJSCC, SING-Zero, and SING-INN under $\rho=0.0052$ and SNR=$5$~dB.}
\label{fig:Visual Results Tiny5}
\end{figure*}
\clearpage

\subsection{Results and Analysis}

Figs. \ref{fig:performance0052} and \ref{fig:performance0013} present the performance comparison of DeepJSCC, trained and tested on the CelebA-HQ dataset, with the proposed two-stage methods SING-Zero and SING-INN, in terms of PSNR and LPIPS as functions of channel SNR (dB). Specifically, Fig. \ref{fig:performance0052} presents results for $\rho=0.0052$, while Fig. \ref{fig:performance0013} focuses on a more constrained channel bandwidth of $\rho=0.0013$. In both cases, higher PSNR indicates better image quality, whereas lower LPIPS reflects superior perceptual similarity between the ground truth and restored images.

For a bandwidth compression of $\rho=0.0052$, both SING-Zero and SING-INN slightly improve PSNR. While SING-Zero achieves a modest enhancement in LPIPS, SING-INN further boosts perceptual quality by explicitly modeling non-linear degradations. Under more extreme degradation conditions, such as $\rho=0.0013$, the fully zero-shot SING-Zero struggles to improve LPIPS due to the difficulty of approximating severe non-linear degradations as linear transformations. In contrast, SING-INN effectively models non-linear degradations, significantly improving LPIPS compared to DeepJSCC and achieving superior perceptual quality. Fig. \ref{fig:Visual Results} provides visual comparisons of reconstructed images alongside their corresponding original and DeepJSCC outputs at $\rho=0.0013$ for SNR=$-1$~dB and SNR=$5$~dB. These results demonstrate that our methods effectively restore perceptual details lost under extreme channel conditions in DeepJSCC.

Compared to \cite{erdemir2022generative}, we place greater emphasis on LPIPS during training of the baseline DeepJSCC, resulting in a better baseline LPIPS relative to \cite{erdemir2022generative}. This increased focus on perceptual metrics further amplifies the challenge of improving performance using generative models. Under this new baseline, StyleGAN-2 based InverseJSCC fails to improve both pixel-level PSNR and perceptual quality represented by LPIPS. To further demonstrate the robustness of our proposed methods and to enable a direct comparison with the latest inverse-problem-based JSCC method, InverseJSCC, we introduced an even more challenging setting under extreme channel conditions. Specifically, we trained the baseline DeepJSCC model on TinyImageNet, a dataset with only 100,000 natural images (less than 1\% of the 14,000,000 images in standard ImageNet) and a resolution of just 64x64. We then tested this model on the same high-resolution 512x512 CelebA-HQ dataset used in previous experiments. This substantial distributional mismatch between training and test data significantly amplifies the difficulty of this inverse problem.

\begin{figure}[!t]\footnotesize %
\centering
\hspace{-0.26cm}
\begin{tabular}{c@{\extracolsep{0em}}c@{\extracolsep{0em}}c}
       
		\includegraphics[width=0.35\textwidth]{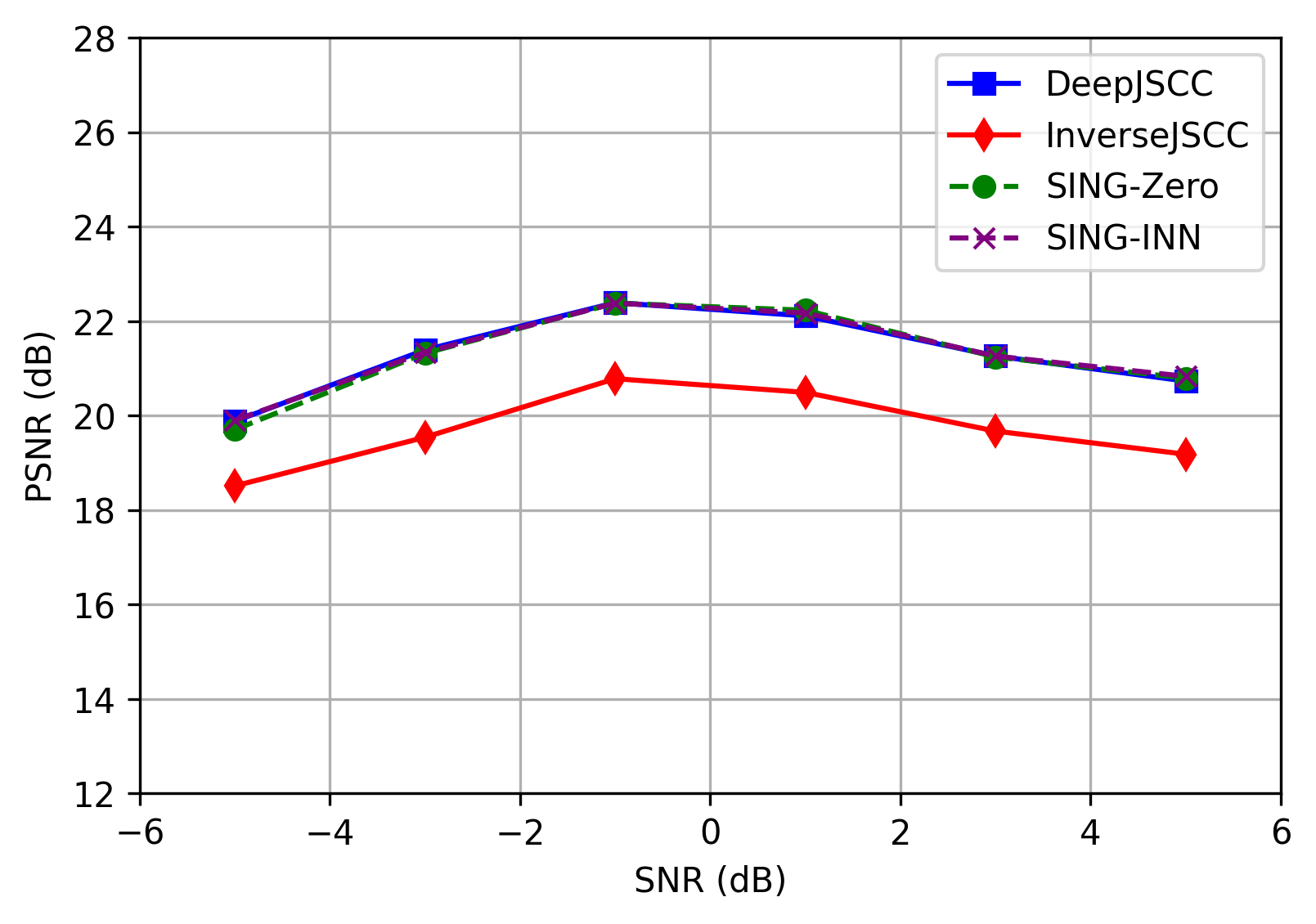}\\
  (a) PSNR versus SNR (higher better)\\
		\includegraphics[width=0.35\textwidth]{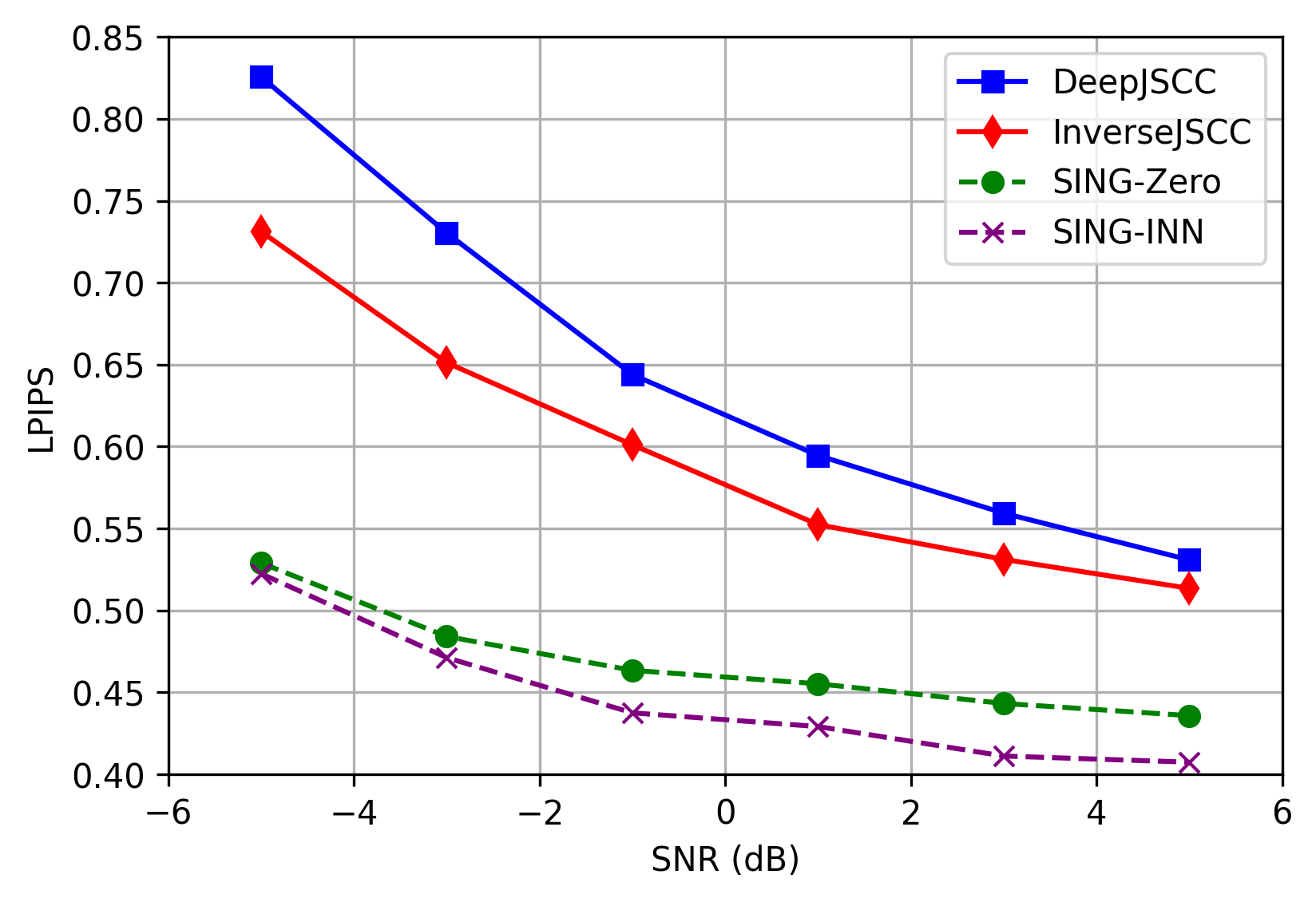}~\\
   (b) LPIPS versus SNR (lower better) \\
	\end{tabular}
    \vspace{-0.1cm} 
	\caption{Performance comparison when DeepJSCC is trained on TinyImageNet in terms of PSNR and LPIPS for $\rho=0.0052$. } 
 \vspace{-0.1cm}
	\label{fig:performance0052_Tiny} 
 \vspace{-0.3cm}
\end{figure}

We compare DeepJSCC, StyleGAN-2-based InverseJSCC, and our proposed SING-Zero and SING-INN methods under the baseline where DeepJSCC was trained on TinyImageNet and tested on CelebA-HQ, as shown in Fig. \ref{fig:performance0052_Tiny}. For fairness, the generative priors—StyleGAN-2 and DDPM—were both trained on the FFHQ dataset. We can observe that while InverseJSCC achieves better LPIPS than DeepJSCC across all channel SNR values, it sacrifices PSNR performance. Our fully zero-shot SING-Zero method significantly outperforms InverseJSCC in LPIPS across all SNR levels while maintaining comparable PSNR performance. The SING-INN approach further enhances perceptual quality, especially at lower SNR values, where improvements in LPIPS over InverseJSCC become increasingly pronounced. Fig. \ref{fig:Visual Results Tiny1} and Fig. \ref{fig:Visual Results Tiny5} provide visual comparisons of reconstructed images when DeepJSCC is trained
on TinyImageNet alongside their corresponding original and DeepJSCC outputs for SNR=${1,5}$ at $\rho=0.0052$.

\section{Conclusion}
\label{sec:conclusion}

In this paper, we introduced SING, a novel two-stage JSCC scheme for wireless image transmission that formulates the reconstruction of received corrupted images as an inverse problem. Depending on whether partial information about the DeepJSCC system and the channel is available, the degradation introduced by the channel and DeepJSCC can either be approximated as a linear transformation or modeled using an INN. Both the linear approximation and INN modeling enable integration of diffusion models into the reconstruction process. SING delivers image reconstructions with perceptual quality surpassing both DeepJSCC and InverseJSCC, particularly in extreme scenarios with very low BCRs and SNRs. Even in more challenging cases where there is a significant distribution mismatch between the training and testing dataset of DeepJSCC, SING demonstrates remarkable performance.

\bibliographystyle{IEEEtran}  
\bibliography{main}

\end{document}